\documentclass[sigconf]{acmart}
\usepackage{float}
\usepackage{stfloats}
\usepackage{graphicx}
\usepackage{flushend}   
\usepackage{microtype}  
\usepackage{subcaption}
\usepackage{caption}
\usepackage{enumitem}
\usepackage{listings}
\usepackage{etoolbox}
\AtBeginDocument{%
  }

\copyrightyear{2026}
\acmYear{2026}
\setcopyright{cc}
\setcctype{by}
\acmConference[IUI '26]{31st International Conference on Intelligent User Interfaces}{March 23--26, 2026}{Paphos, Cyprus}
\acmBooktitle{31st International Conference on Intelligent User Interfaces (IUI '26), March 23--26, 2026, Paphos, Cyprus}
\acmPrice{}
\acmDOI{10.1145/3742413.3789125}
\acmISBN{979-8-4007-1984-4/2026/03}

\begin{document}

\title{GuideAI: A Real-time Personalized Learning Solution with Adaptive Interventions}


\author{Ananya Shukla}
\authornote{These authors contributed equally to the work.}
\affiliation{%
  \institution{HTI Lab, Plaksha University}
  \city{Mohali}
  \country{India}
}
\email{ananya.shukla@plaksha.edu.in}

\author{Chaitanya Modi}
\authornotemark[1]
\affiliation{%
  \institution{HTI Lab, Plaksha University}
  \city{Mohali}
  \country{India}
}
\email{chaitanya.modi@plaksha.edu.in}

\author{Satvik Bajpai}
\authornotemark[1]
\affiliation{%
  \institution{HTI Lab, Plaksha University}
  \city{Mohali}
  \country{India}
}
\email{satvik.bajpai@plaksha.edu.in}

\author{Siddharth Siddharth}
\affiliation{%
  \institution{HTI Lab, Plaksha University}
  \city{Mohali}
  \country{India}
}
\email{siddharth.s@plaksha.edu.in}

\renewcommand{\shortauthors}{Shukla et al.}

\begin{teaserfigure}
  \includegraphics[width=\textwidth]{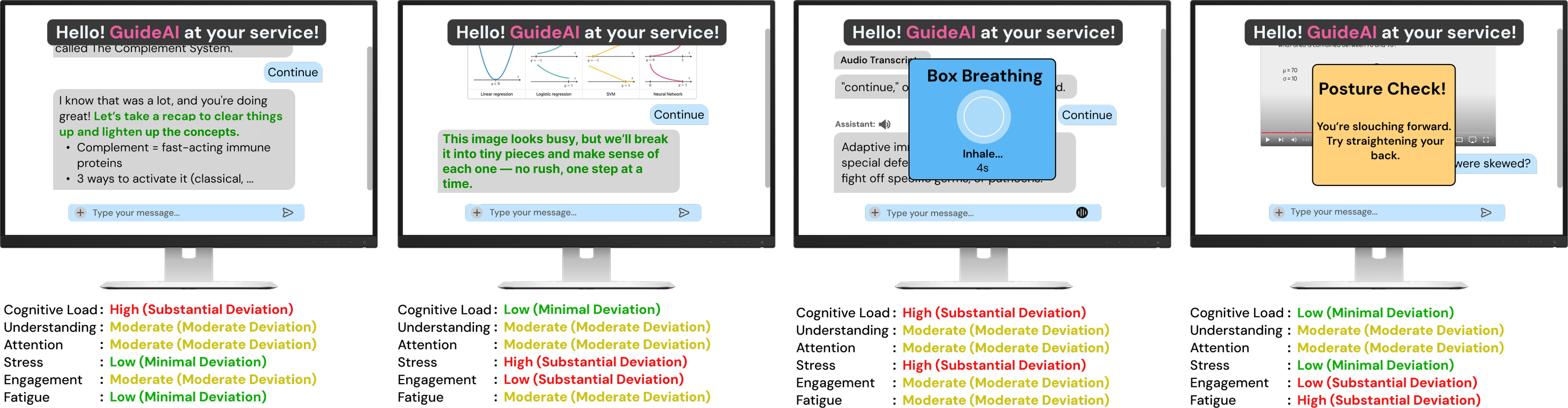}
  \caption{GuideAI's Comprehensive Adaptive Learning Interface. The image illustrates the multi-modal, biosensor-driven approach of the personalized system across different learning modes---text-based, image-based, audio-based and video-based. Each panel demonstrates real-time adaptation to the learner's cognitive state, featuring dynamic interventions such as cognitive load management, stress reduction techniques (box breathing), and posture correction.}
\Description{Four computer monitors arranged side-by-side displaying the GuideAI interface in different learning modes. 
1) The first screen shows a text-based chat interface with a sidebar indicating 'High Cognitive Load'. 
2) The second screen displays an image-based lesson with a graph and an overlay explaining the concept. 
3) The third screen shows an audio transcript with a large blue pop-up overlay titled 'Box Breathing' with an animation circle. 
4) The fourth screen shows a video player with an orange pop-up warning titled 'Posture Check!' advising the user to straighten their back. 
Below each monitor are real-time status indicators for Cognitive Load, Understanding, Attention, Stress, Engagement, and Fatigue, showing varying levels of deviation.}
  \label{fig:teaser}
\end{teaserfigure}

\begin{abstract}
Large Language Models (LLMs) have emerged as powerful learning tools, but they lack awareness of learners' cognitive and physiological states, limiting their adaptability to the user's learning style. Contemporary learning techniques primarily focus on structured learning paths, knowledge tracing, and generic adaptive testing but fail to address real-time learning challenges driven by cognitive load, attention fluctuations, and engagement levels. Building on findings from a formative user study ($N=66$), we introduce GuideAI, a multi-modal framework that enhances LLM-driven learning by integrating real-time biosensory feedback including eye gaze tracking, heart rate variability, posture detection, and digital note-taking behavior. GuideAI dynamically adapts learning content and pacing through cognitive optimizations (adjusting complexity based on learning progress markers), physiological interventions (breathing guidance and posture correction), and attention-aware strategies (redirecting focus using gaze analysis). Additionally,  GuideAI supports diverse learning modalities, including text-based, image-based, audio-based, and video-based instruction, across varied knowledge domains. A preliminary study (N = 25) assessed GuideAI’s impact on knowledge retention and cognitive load through standardized assessments. The results show statistically significant improvements in both problem-solving capability and recall-based knowledge assessments. Participants also experienced notable reductions in key NASA-TLX measures including mental demand, frustration levels, and effort, while simultaneously reporting enhanced perceived performance. These findings demonstrate GuideAI's potential to bridge the gap between current LLM-based learning systems and individualized learner needs, paving the way for adaptive, cognition-aware education at scale.
\end{abstract}



\maketitle
\sloppy 

\section{Introduction}
The rapid evolution of Large Language Models (LLMs) has transformed domains such as content generation, scientific discovery, and medicine. Their ability to process and generate natural language has opened new frontiers in Human–Computer Interaction (HCI) \cite{Bommasani2022}. In education, integrating LLMs with learning technologies shows particular promise: they can provide instant feedback, scaffold explanations, and adapt responses to learner inputs \cite{Sanchez2024, Molina2024, Gupta2025}.

Recent studies have demonstrated that LLMs have proven effective as tutoring systems, capable of answering conceptual questions, explaining complex topics, supporting problem-solving, and promoting meta-cognitive reflection on knowledge gaps \cite{SiyangLiu2024}. However, most current systems remain text-centric and cognitively myopic, offering limited personalization and no awareness of learners’ real-time cognitive-affective states \cite{Alhafni2024}.

Learning science research emphasizes that effective learning depends on context and fluctuates with attention, cognitive load, affect, and physiological well-being \cite{Madsen2021}. Ideally, intelligent tutors would infer these states from multi-modal signals such as gaze, heart rate variability (HRV), posture, or interaction traces, and adapt their strategies accordingly. Yet current LLM tutors remain reactive, relying only on textual cues like correctness, verbosity, or sentiment. This single-channel approach treats learning as static text-exchange, overlooking the dynamic interplay of attention, uncertainty, and affect that drives genuine learning.

\sloppy
This gap can be conceptualized along four critical dimensions that constrain current LLM-based learning systems:

\begin{enumerate}[leftmargin=*]

\item \textbf{Representation Gap — Textual Proxies for Cognition:}
Current systems assess learner's understanding mainly through surface-level text features such as accuracy, verbosity, or sentiment. These overlook deeper indicators of learning—like attention, uncertainty, and cognitive effort—that appear in physiological and behavioral signals rather than language \cite{D’Mello2010, D'Mello2015}.

\item \textbf{Temporal Gap — Lack of Real-time Adaptivity:}
Most LLM tutors follow a turn-based dialogue model, updating only after learner's input. This ignores the continuous dynamics of cognition and emotion, limiting timely interventions that could sustain focus or prevent overload \cite{Dissanayake2025}.

\item \textbf{Inference Gap — Absence of Multi-modal State Modeling:}
Existing systems seldom use multi-modal data such as gaze, HRV, or posture to infer learner's real-time cognitive-affective states. Without these cues, personalization remains context-blind to engagement, fatigue, or frustration \cite{Zhu2025}.

\item \textbf{Intervention Gap — Lack of Adaptive, Flow-preserving Support:}
Even when learner's states are inferred, most systems lack adaptive frameworks that decide when and how to intervene. They rely on static prompts or fixed thresholds instead of context-aware, closed-loop strategies that preserve cognitive flow and support metacognitive control \cite{Dissanayake2025, Lim2025}.
\end{enumerate}

To address these limitations, we first conducted a formative assessment ($N=66$) to identify learner needs. Guided by these results, this study addresses these gaps by introducing GuideAI, a novel biosensor-augmented personalized LLM system across diverse learning modalities. Our work makes the following contributions:

\begin{enumerate}[leftmargin=*]
\item \textbf{Integrated Biosensory Framework:}
We present a unified behavioral-physiological sensing framework that fuses real-time multi-modal signals such as gaze, HRV, posture, and interaction traces within an LLM-based learning environment. This enables continuous inference of cognitive-affective states including engagement, cognitive load, and fatigue, forming the basis for context-aware and tone-sensitive feedback.

\item \textbf{Adaptive Intervention Strategies:}
We design a closed-loop intervention model that dynamically determines when and how to respond based on inferred learner states. GuideAI delivers tiered interventions at the micro-level (content pacing and complexity), meso-level (focus and affect regulation), and macro-level (metacognitive guidance) to sustain flow and promote self-regulation

\item \textbf{Cross-modal Interface and Empirical Validation:}
We implement GuideAI in a multi-modal environment across text, image, audio, and video formats, and evaluate it through a mixed-method study (N = 25) combining biosensor metrics with self-reported measures. Results demonstrate statistically significant improvements in retention, understanding, and cognitive load management compared to non-personalized LLM baselines.
\end{enumerate}

By creating systems that can sense and respond to learners' cognitive and physiological states in real-time, GuideAI bridges the gap between current LLM-based learning systems and the individualized needs of learners, paving the way for adaptive, cognition-aware education at scale. GuideAI's codebase will be released as open source.

\section{Related Work}
\subsection{Physiological and Cognitive States as Determinants of Learning Efficiency}
Building upon the need for personalization, recent research emphasizes that effective learning is not only a cognitive process but also deeply influenced by physiological states. Commins et al. \cite{Commins2018} demonstrated that physical well-being directly modulates cognitive functions essential for effective information processing and knowledge acquisition. Suboptimal physiological states, particularly poor posture, have been linked to increased cognitive load and diminished learning efficiency \cite{Shi2025}. Research on cognitive load theory has established that when mental processing resources are overwhelmed, learning efficiency decreases substantially, highlighting the necessity of monitoring and managing cognitive demands during educational activities \cite{Commins2018, Rodemer2023, Resnick2017}.

\subsection{Biosensors as Biomarkers for Personalized and Adaptive Learning}
Advances in multi-modal biosensing now enable non-intrusive monitoring of learners’ cognitive and emotional states in real time. Pupillometry, for instance, provides reliable indices of mental effort and attentional load, with pupil dilation correlating strongly with task difficulty and processing demand \cite{D'Mello2012, Rodemer2023}. Heart rate variability (HRV) serves as a biomarker of stress and engagement by capturing autonomic regulation \cite{Nicolini2024, Ahmed2025}, while posture dynamics and micro-movements indicate fatigue or disengagement \cite{Jayasinghe2023, Shi2025}. These physiological and behavioral signals offer complementary perspectives on the learner’s internal state however, existing approaches typically use these signals in isolation for offline analytics or post-hoc feedback rather than integrating them into closed-loop adaptive systems \cite{D'Mello2015}

\subsection{From Rule-based Systems to LLMs: Toward Multi-modal Personalization in Education}

The integration of LLMs into educational technologies has revolutionized learning through advanced natural language understanding and generation capabilities \cite{Chu2025, Khanmigo2025, ChatGPTEdu2024}. Contemporary implementations leverage LLMs' sophisticated reasoning processes to adapt instructional approaches based on inferred knowledge states and learning preferences \cite{ChenDing2024, Zhang2024}, however fail to address the temporal and inferential gaps of adaptive support: feedback timing, intervention intensity, and context awareness. The combination of LLMs with wearable biosensors represents a particularly promising direction, with prior studies demonstrating content adaptation and limited learner modeling through physiological data integration \cite{Ahmed2025, Fang2024, Ferrara2024}. These models' multi-modal processing capabilities facilitate the integration of diverse data sources, including textual, visual, physiological, and behavioral metrics, enabling a more comprehensive understanding of learners' states \cite{Xu2024, Zhang2024}.

Despite these advances, most current LLM tutoring systems remain reactive, inferring progress only from textual responses rather than multi-modal learner-state cues. They fail to address the temporal and inferential gaps of adaptive support: feedback timing, intervention intensity, and context awareness \cite{Zhang2020, Nicolini2024}. While prior frameworks have explored content adaptation and limited learner modeling \cite{Laak2025, Ahmed2025, Zhaoli_Zhang2020}, none combine real-time multi-modal biosensing, cognitive flow-aligned adaptation, and meta-cognitive scaffolding within a unified architecture.
GuideAI contributes to this trajectory by operationalizing a context-aware, biosensor-augmented LLM framework that performs continuous learner-state inference and adaptive intervention at multiple temporal and cognitive scales, extending prior work in both educational AI and intelligent user interfaces.

\section{Formative Assessment}
To ground our system design in authentic learner needs, we conducted a formative study to translate broad challenges, limited personalization and lack of real-time adaptation into actionable, user-centered design goals. The assessment captured learners’ AI usage patterns, limitations of current tools, and expectations for adaptive interventions.

\subsection{Pre-study Survey}
A total of 66 participants (mean age = 24.3, range = 19-42) completed an online survey. The sample included undergraduate and graduate students, educators, and professionals in learning-focused roles (63.3\% male, 36.7\% female). The survey examined participants’ AI usage frequency, perceived utility, preferred learning modalities, and desired personalization features.

\begin{figure}[ht]
    \centering
    \includegraphics[width=\columnwidth]{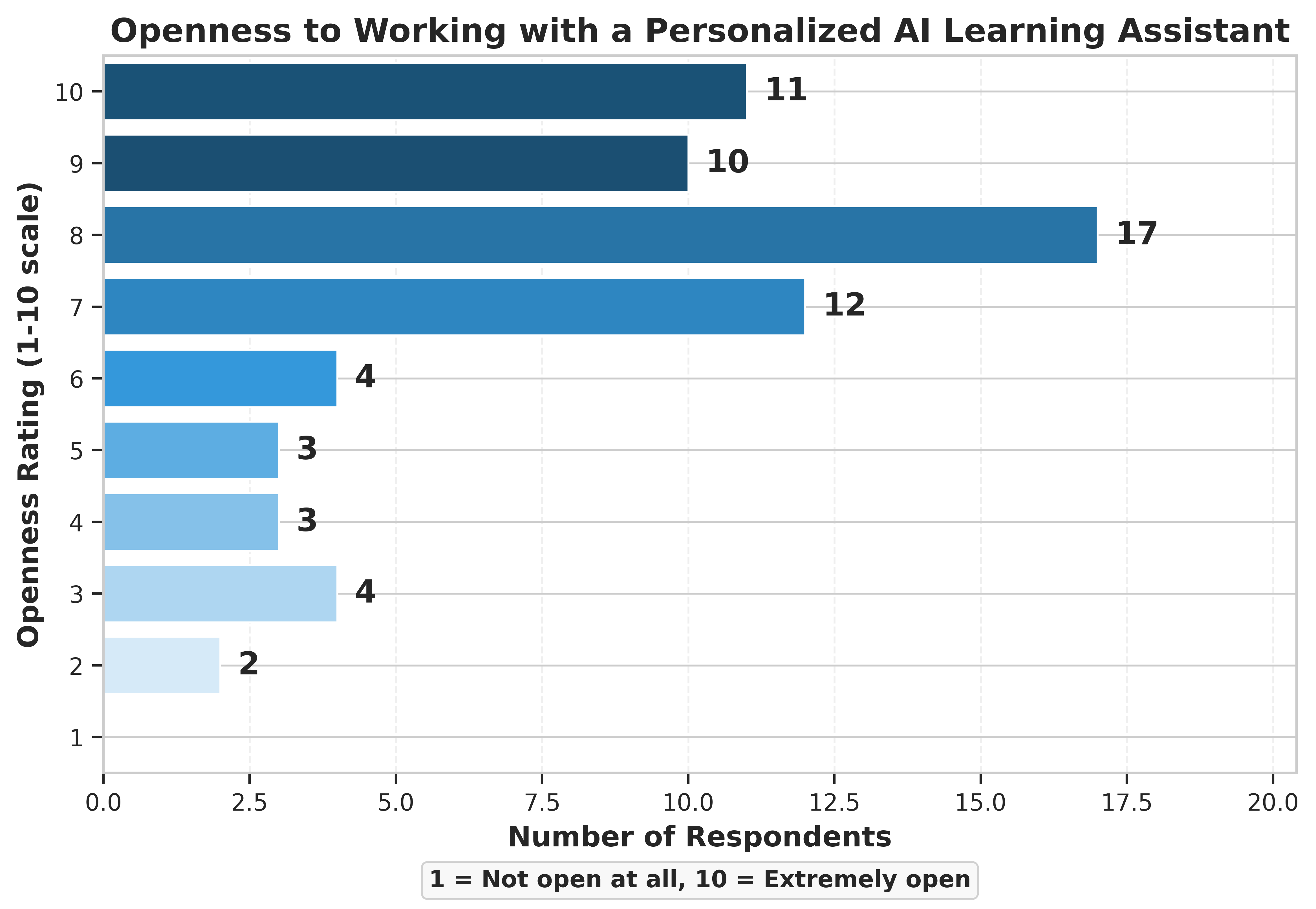}
    \caption{Participant openness to working with a personalized AI learning assistant on a scale from 1-10. The distribution shows strong receptivity, with 50 out of 66 participants (75.7\%) rating their openness at 7 or higher, and a median rating of 8.}
    \label{fig:ai_openness}
    \Description{Horizontal bar chart titled 'Openness to Working with a Personalized AI Learning Assistant' plotted on a scale from 1 to 10. The distribution is heavily skewed towards high openness. 
The specific counts are: Rating 10 (17 respondents), Rating 9 (12 respondents), Rating 8 (11 respondents), Rating 7 (10 respondents), Rating 6 (4 respondents), Rating 5 (4 respondents), Rating 4 (4 respondents), Rating 3 (4 respondents), Rating 2 (0 respondents). 
The chart indicates that 75.7\% of participants rated their openness at 7 or higher.}
\end{figure}

\subsubsection{AI Usage and Perception}
Most participants (55.0\%) used AI tools daily and 38.3\% several times per week. Nearly all respondents (96.6\%) found AI systems at least somewhat useful for learning, with 40.7\% rating them extremely useful. \textit{ChatGPT} was the most used platform (96.7\%), followed by \textit{Claude} (61.7\%) and \textit{Gemini} (30.0\%). Participants viewed AI primarily as an on-demand explainer for complex problems, rather than a continuous learning companion, indicating opportunities for more proactive, context-aware assistance.

\subsubsection{Learning Modality Preferences}
Text-based materials remained the most preferred learning modality (98.3\%), though many participants expressed interest in video (53.3\%), image-based (51.7\%), and audio (44.6\%) content. Text was valued for self-paced control, while visual and auditory formats aided understanding of spatial or procedural concepts. As one participant noted, “\textit{Text lets me learn at my own speed, but diagrams and videos help me grasp complex ideas much faster}” (P7). This preference underscores the importance of multi-modal learning environments that flexibly align with learners’ cognitive and perceptual needs.

\subsubsection{AI Learning Tools: Usage Patterns and Limitations}
Participants primarily used AI tools to explain difficult concepts (91.7\%), interpret code or mathematical problems (78.3\%), and summarize readings (73.3\%). These findings indicate that learners rely on AI systems during moments of cognitive friction rather than routine study.

However, participants identified several persistent limitations. About one-third (31.7\%) cited a lack of personalization or adaptation to their learning style as a major shortcoming. Others noted excessive textual explanations (33.3\%) or missing step-by-step reasoning (25.0\%). As participants described, systems often “\textit{keep explaining at the same level even when I’m lost}” (P12) or “\textit{can't tell when an explanation isn't landing}” (P15). Nearly half (42.3\%) reported abandoning AI assistance when it failed to adjust to their confusion. These limitations highlight the absence of real-time learner-state awareness—precisely the challenge GuideAI addresses through multi-modal biosensing and adaptive feedback.

\begin{figure*}[!htp] 
    \centering
    \includegraphics[width=0.85\textwidth]{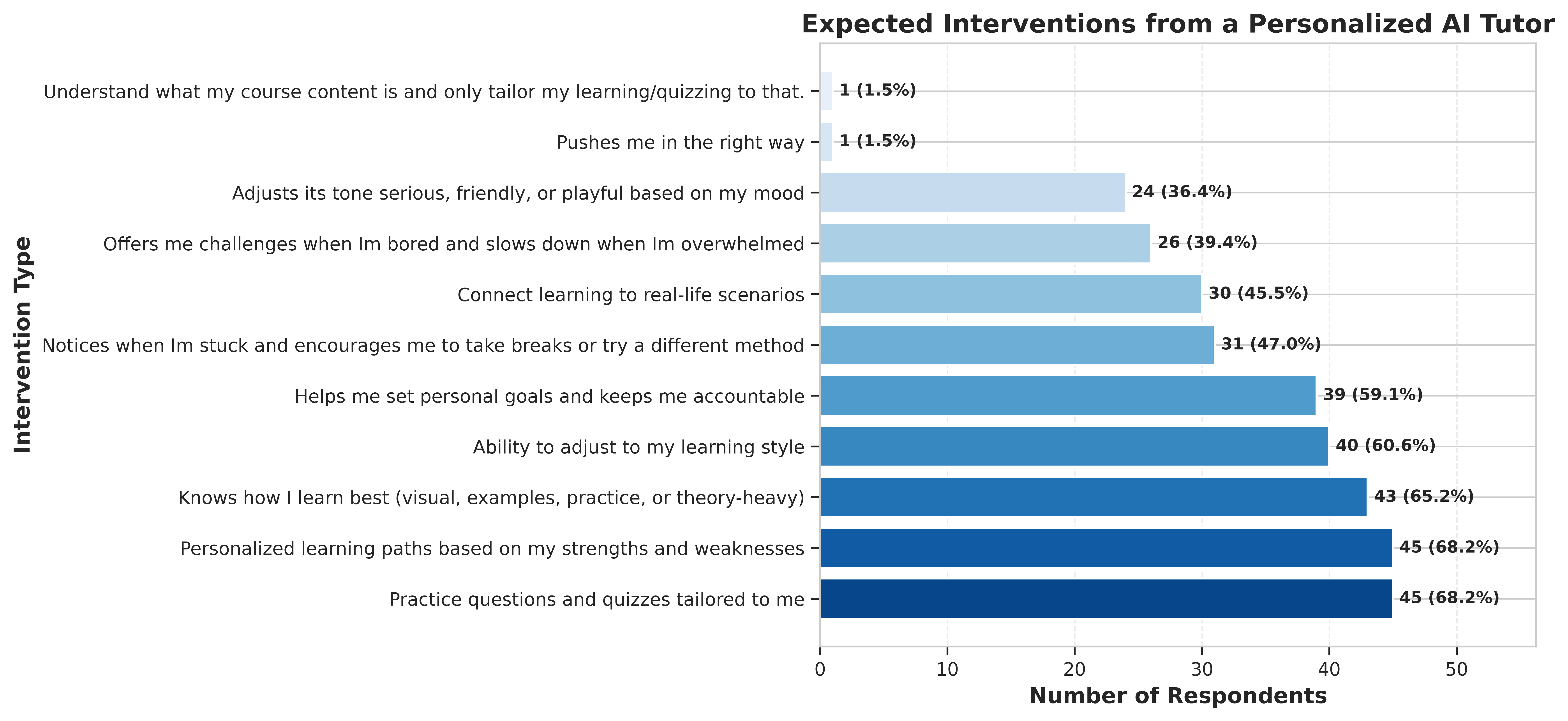} 
\caption{Desired interventions from personalized AI (N = 66). Top priorities: personalized learning paths (68.2\%), tailored practice questions (68.2\%), and learning style adaptation (65.2\%). Nearly half (47.0\%) wanted break reminders and alternative methods. Multiple selections allowed.}
    \label{fig:desired_interventions}
    \Description{Horizontal bar chart ranking 'Expected Interventions from a Personalized AI Tutor'. 
The top three most requested features are tied or nearly tied: 'Practice questions and quizzes tailored to me' (68.2\%, 45 respondents) and 'Personalized learning paths based on strengths and weaknesses' (68.2\%, 45 respondents), followed by 'Knows how I learn best' (65.2\%, 43 respondents). 
Other features include 'Ability to adjust to my learning style' (60.6\%), 'Helps me set personal goals' (59.1\%), 'Notices when I am stuck' (47.0\%), 'Connect learning to real-life scenarios' (45.5\%), 'Offers challenges when bored' (39.4\%), and 'Adjusts tone' (36.4\%). The least requested features are 'Pushes me in the right way' (1.5\%) and 'Only tailor my learning/quizzing' (1.5\%).}
\end{figure*}

\subsubsection{Desired Interventions}

When asked about desired capabilities in AI learning assistants, participants expressed strong interest in adaptive interventions (Figure~\ref{fig:ai_openness}). Top-ranked features included personalized learning paths (68.3\%), targeted practice questions (66.7\%), and adaptive recognition of learning styles (65.0\%) (Figure~\ref{fig:desired_interventions}). Over half (50.0\%) wanted systems capable of detecting struggle and suggesting breaks or alternative strategies. These findings directly informed GuideAI’s design goals—developing a system that dynamically infers cognitive-affective state, modulates support timing and intensity, and maintains learner engagement across modalities.

\subsection{Design Implications}
The formative assessment yielded four key design implications that guided GuideAI's development:

\begin{enumerate}[leftmargin=*]
    \item \textbf{Multi-modal Learning Support}: Significant interest in image-based, audio-based and video-based learning suggests the need for systems that seamlessly integrate and adapt across modalities.
    
    \item \textbf{Real-time Adaptation}: Users desire systems that can detect learning states and dynamically adjust content and pace accordingly.

    \item \textbf{Physiological Awareness}: Strong interest in systems that can detect struggles and suggest interventions indicates the value of incorporating biosensory data.

    \item \textbf{Personalized Learning Paths}: Respondents prioritized individualized content sequencing based on their unique strengths and weaknesses, supporting our biosensory-informed personalization approach.
\end{enumerate}

These findings highlighted the gap between current AI learning systems and learner needs, directly informing our GuideAI system design that integrates biosensory data for responsive, personalized learning across multiple modalities.

\section{GuideAI Solution}

\begin{figure*}[!htp] 
    \centering
    \includegraphics[width=\textwidth]{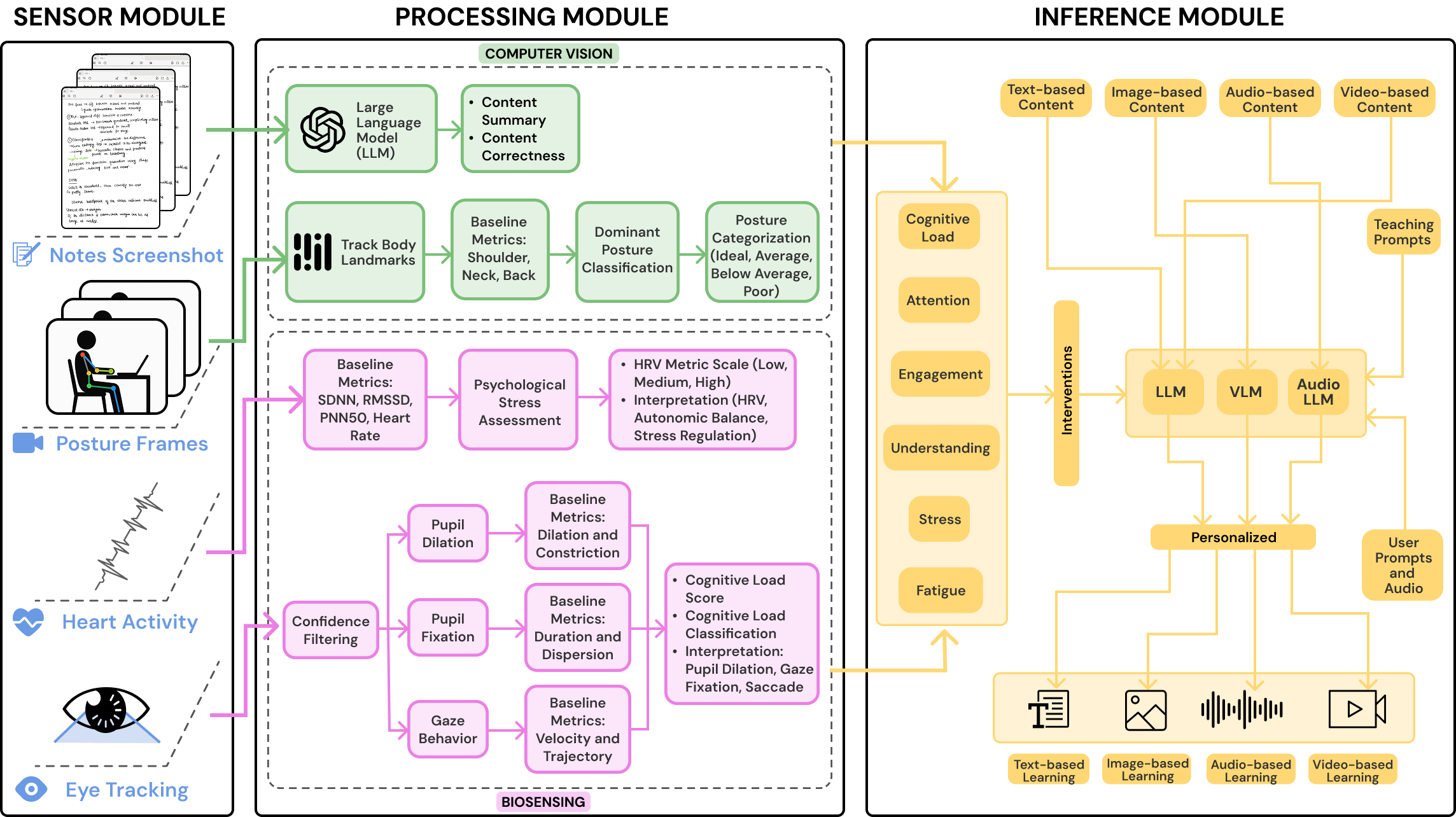} 
\caption{GuideAI System Architecture with three modules: Sensor Module (left) collects eye tracking, heart activity, posture, and notes data. Processing Module (center) transforms raw signals into physiological and behavioral metrics. Inference Module (right) integrates these to assess six cognitive dimensions (cognitive load, attention, engagement, understanding, stress, fatigue) and deliver personalized interventions across learning modalities (text, image, audio, video).}
    \label{fig:architecture}
    \Description{A detailed system architecture diagram divided into three main columns: Sensor Module, Processing Module, and Inference Module.
1) The Sensor Module (left) shows inputs from Notes Screenshots, Posture Frames (camera), Heart Activity (wearable), and Eye Tracking (glasses).
2) The Processing Module (center) shows parallel pipelines: A Computer Vision pipeline processing posture landmarks and using an LLM to summarize notes; and a Biosensing pipeline processing Heart Rate (HRV metrics like SDNN, RMSSD) and Eye Tracking data (Pupil dilation, blink rate, gaze velocity).
3) The Inference Module (right) aggregates these signals to compute six cognitive states: Cognitive Load, Attention, Engagement, Understanding, Stress, and Fatigue. These states feed into an 'Interventions' engine that dynamically adjusts the output via an LLM, VLM, or Audio LLM to generate personalized Text, Image, Audio, or Video content.}
\end{figure*}

Following our formative studies, we developed GuideAI, a personalized system that identifies challenges and support opportunities based on multi-modal cognitive and behavioral cues.

\subsection{Overview}
The GuideAI pipeline (Figure~\ref{fig:architecture}) comprises three core components: 
(1) A sensor module, that includes an external posture camera, pupil-tracker, wrist-worn wearable, and note-taking device to monitor a learner's physiological and behavioral state during digital learning sessions. (2) A processing module that analyzes the captured data streams to infer the learner’s engagement, cognitive workload, and affective state. (3) A reasoning and inference module that integrates contextual understanding with pedagogical strategies to suggest timely interventions.\\
We implemented GuideAI as a web-based platform that supports real-time learning and assessment of study behavior. The platform supports cross-device compatibility, enabling data input from tablets, mobile devices, and desktop systems.

\subsection{Data Capture}
\begin{figure}[ht]
    \centering
    \includegraphics[width=\columnwidth]{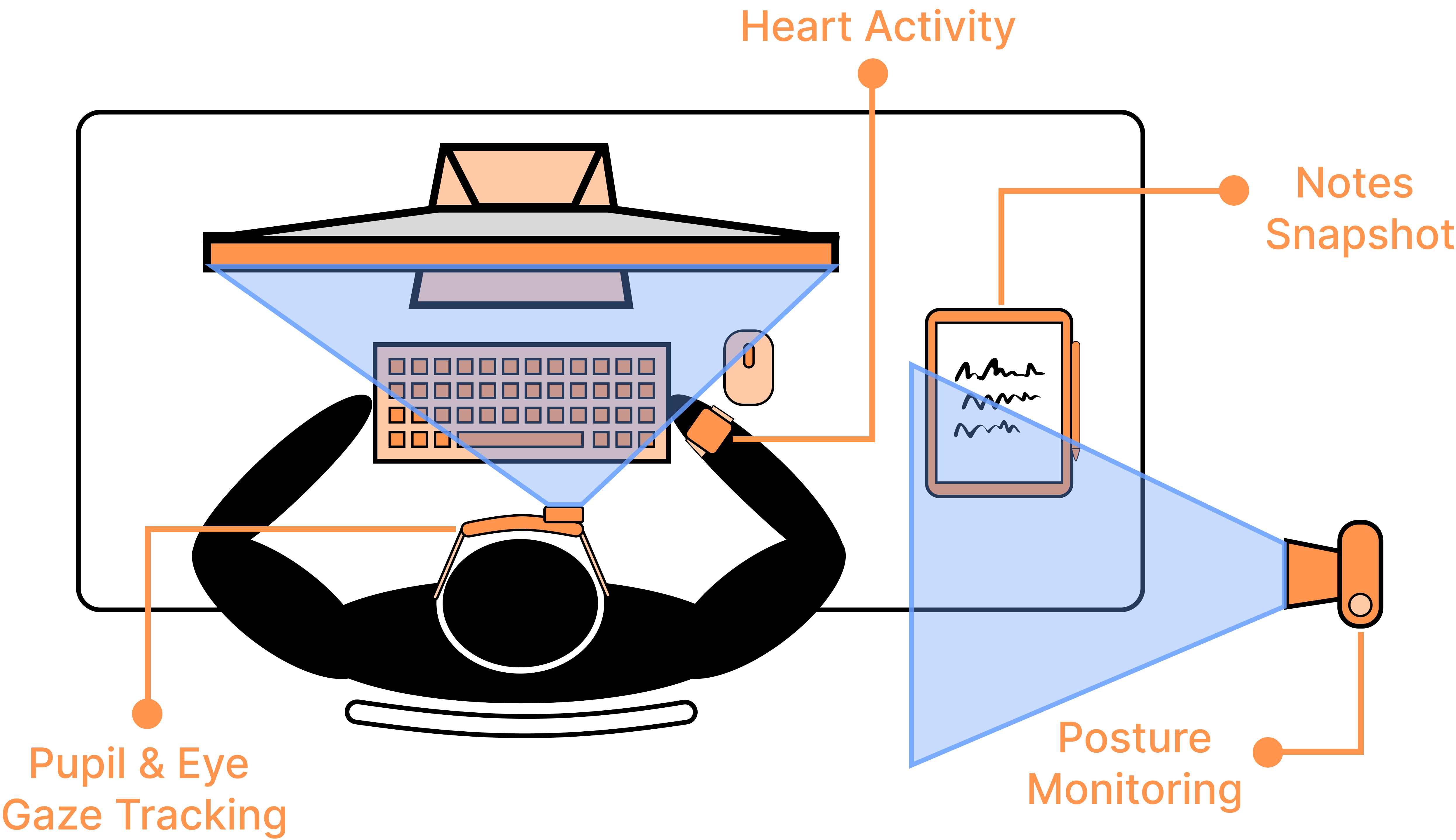}
    \caption{Multi-modal biosensory setup integrating eye tracking, heart rate monitoring, note-taking analysis, and posture detection.}
    \label{fig:sensor_setup}
    \Description{A schematic top-down illustration of the user study setup. A user is seated at a desk with a computer monitor and keyboard. 
Lines point to specific hardware components: 
1) 'Pupil & Eye Gaze Tracking' glasses worn by the user. 
2) 'Heart Activity' sensor on the user's wrist. 
3) 'Notes Snapshot' tablet (iPad) placed to the right of the keyboard. 
4) 'Posture Monitoring' camera positioned to the side of the user capturing their profile.}
\end{figure}

In our experimental setup (Figure~\ref{fig:sensor_setup}), GuideAI uses a multi-modal sensor configuration to collect real-time data from the learner. We use the Pupil Core eye tracker from Pupil Labs \cite{Kassner2014} to capture pupillometry data, fixations, and saccadic movements at 60 fps, providing information about visual attention, cognitive load, and information processing patterns. The Max-Health-Band from Maxim Integrated \cite{Ferrati2023} captures heart rate (HR) and heart rate variability  (HRV) at 200 Hz, enabling assessment of autonomic nervous system activity and stress levels.\\ 
An external posture camera captures upper body positioning, while iPad screen recordings provide observational data on note-taking behaviors. To manage time-synchronized data flow across sensors, we use the Lab Streaming Layer (LSL) framework \cite{Kothe2024}. LSL ensures consistent time-stamping and low-latency data integration, enabling real-time fusion of biosignals, posture analysis, and task interaction data for constructing a temporally coherent model of learners' attention, affect, and cognitive effort.

\subsection{Processing Module}
\subsubsection{Biosensors-based Pipeline}
To infer users’ cognitive and affective states in real-time, our system integrates physiological signals from pupillometry and HRV, extracted via biosensor processing pipelines. The eye-tracking module tracks and analyzes pupil dilation, saccades, and fixations, all of which have been demonstrated to reflect cognitive load, arousal, and visual attention dynamics. Fixation durations correlate with working memory load and attentional allocation during scene perception and memory integration tasks \cite{Goldberg1991, Brockmole2005, Loh2023}. By employing robust fixation-detection algorithms \cite{Salvucci2000, Munn2008}, the system parses gaze data into real-time attentional patterns. Saccadic movements are also analyzed, offering rapid temporal markers of stimulus-driven attentional shifts \cite{Fischer1993, Fischer1987}. These signals are instrumental in detecting user interest, task engagement, and transitions between visual targets.\\
To assess autonomic nervous system activity, we implement a HRV processor that analyzes data from a wearable physiological sensor. The heart activity signal is sampled at 200 Hz and processed in real-time. The system extracts standard time-domain HRV features, focusing on pNN50, RMSSD, and SDNN metrics \cite{Ewing1984, DeGiorgio2010, Wang2012}. The pNN50 value measures the percentage of successive RR intervals differing by more than 50 milliseconds, with thresholds indicating stress levels: <20 (high), 20-50 (moderate), >50 (low stress) \cite{Salahuddin2007, Ziegler2004}. These HRV metrics are contextualized against individual baseline recordings to account for inter-individual differences in resting autonomic tone.

\paragraph{Raw Signal Processing and Feature Extraction}
Raw biosignals are processed in real-time using modality-specific pipelines prior to state inference. For pupillometry, raw pupil diameter streams sampled at 60 Hz are first smoothed using a rolling median filter to remove blink artifacts, followed by baseline normalization. Fixations are detected using a velocity-threshold identification (I-VT) method, where gaze velocity $v_t$ is computed as:

\begin{equation}
v_t = \frac{\sqrt{(\Delta x_t)^2 + (\Delta y_t)^2}}{\Delta t}
\end{equation}
\\
where $x_t$ and $y_t$ represent the horizontal and vertical gaze positions at time $t$, respectively. $\Delta x_t = x_t - x_{t-1}$ and $\Delta y_t = y_t - y_{t-1}$ are the gaze displacements along each axis, and $\Delta t$ is the sampling interval between successive measurements. Samples with $v_t < \theta_v$ are classified as fixations, enabling computation of fixation duration and dispersion metrics.
\\
For cardiovascular signals, RR intervals extracted from the wearable sensor are processed over sliding windows to compute standard HRV features including RMSSD, SDNN, and pNN50. These features are normalized against participant baselines to account for inter-individual physiological variability.

\subsubsection{Vision-based Pipeline}
GuideAI tracks learner engagement through posture analysis and note-taking behavior. For posture monitoring, we use MediaPipe \cite{Lugaresi2019} to analyze shoulder positioning, neck alignment, and back straightness, categorizing posture into four states: ideal (90-100\%), average (75-89\%), below average (60-74\%), and poor ($<$60\%).

For note-taking analysis, an iPad M1 captures screenshots at 60-second intervals. These are processed using OpenAI's \texttt{gpt-4o} model to analyze handwritten notes, providing qualitative feedback and a correctness score (0-1). This integration captures both physical engagement and cognitive processing reflected in learning artifacts.

\subsection{Personalized Intervention Module}

\begin{figure*}[!htp]
    \centering
    \begin{tabular}{cc}
        \includegraphics[width=0.47\textwidth]{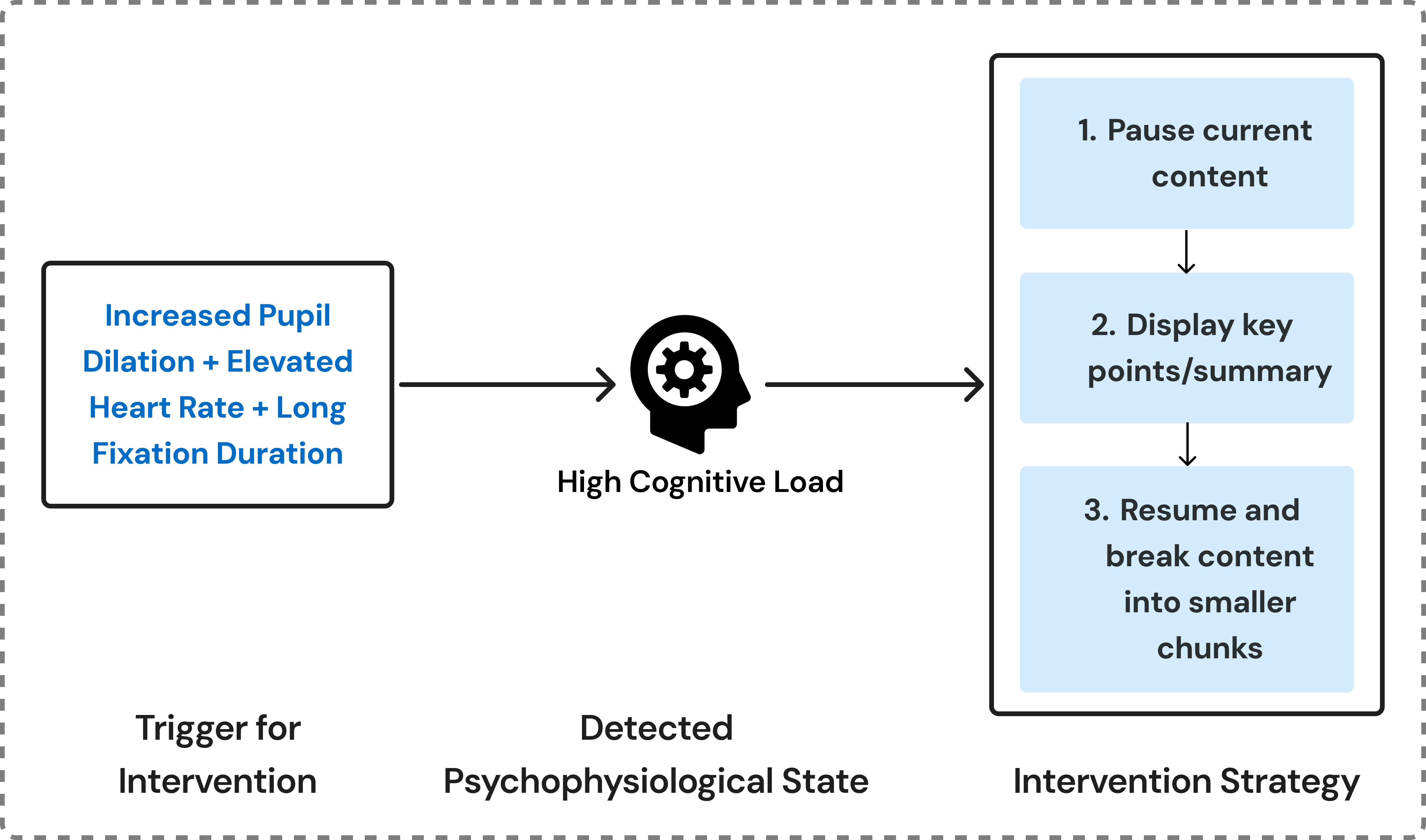} &
        \includegraphics[width=0.47\textwidth]{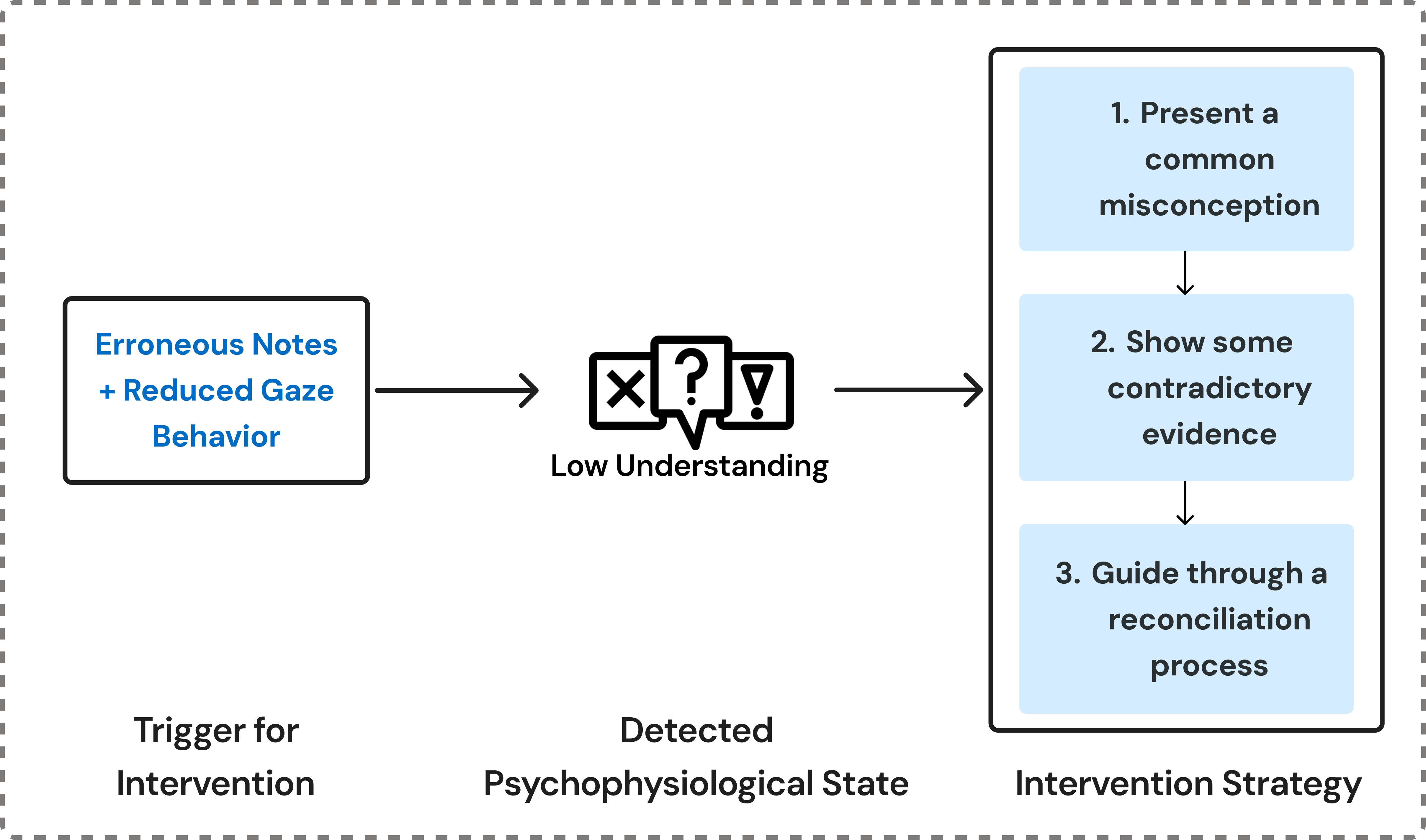} \\
        \small{(a) High Cognitive Load Detection} & 
        \small{(b) Low Understanding Intervention} \\
        \\
        \includegraphics[width=0.47\textwidth]{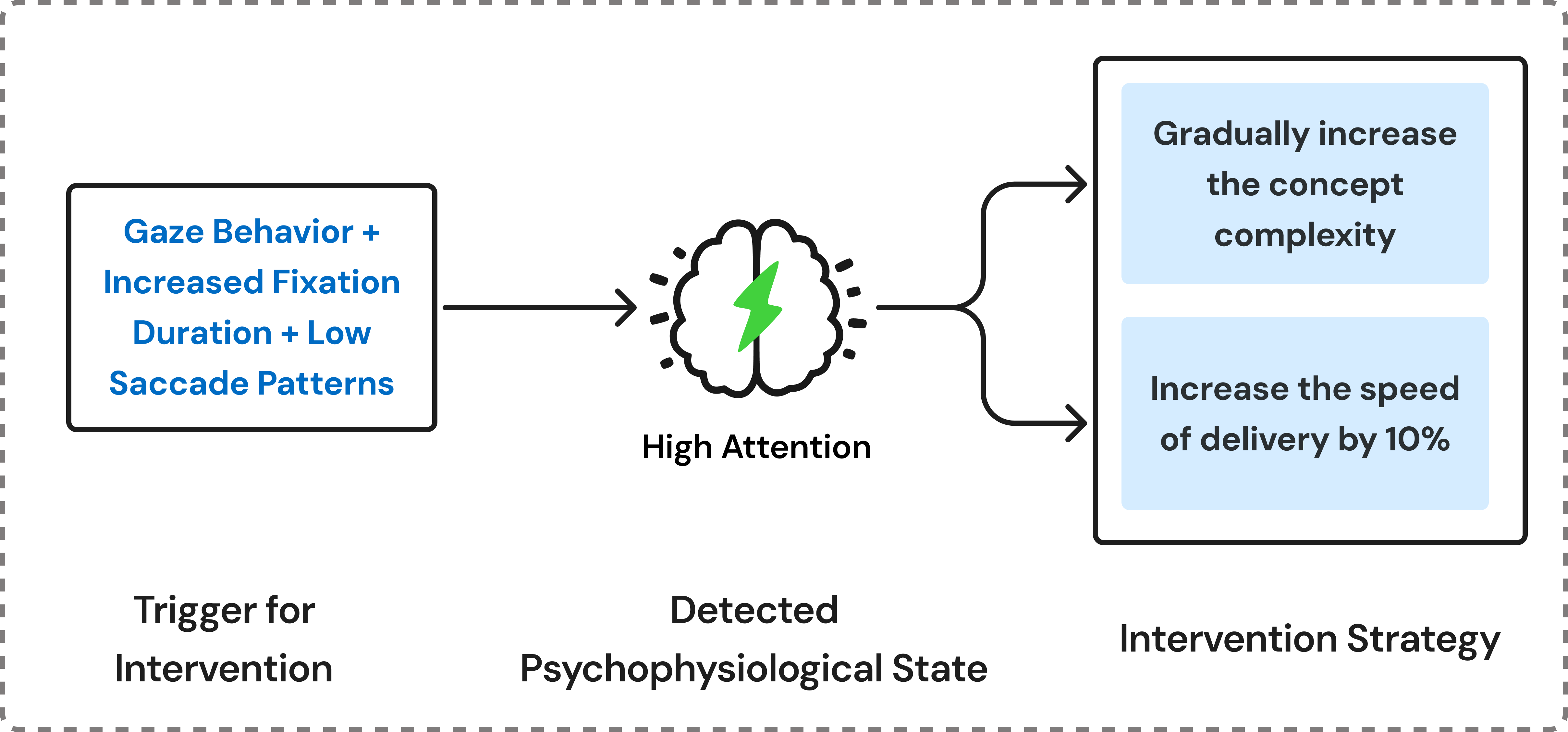} &
        \includegraphics[width=0.47\textwidth]{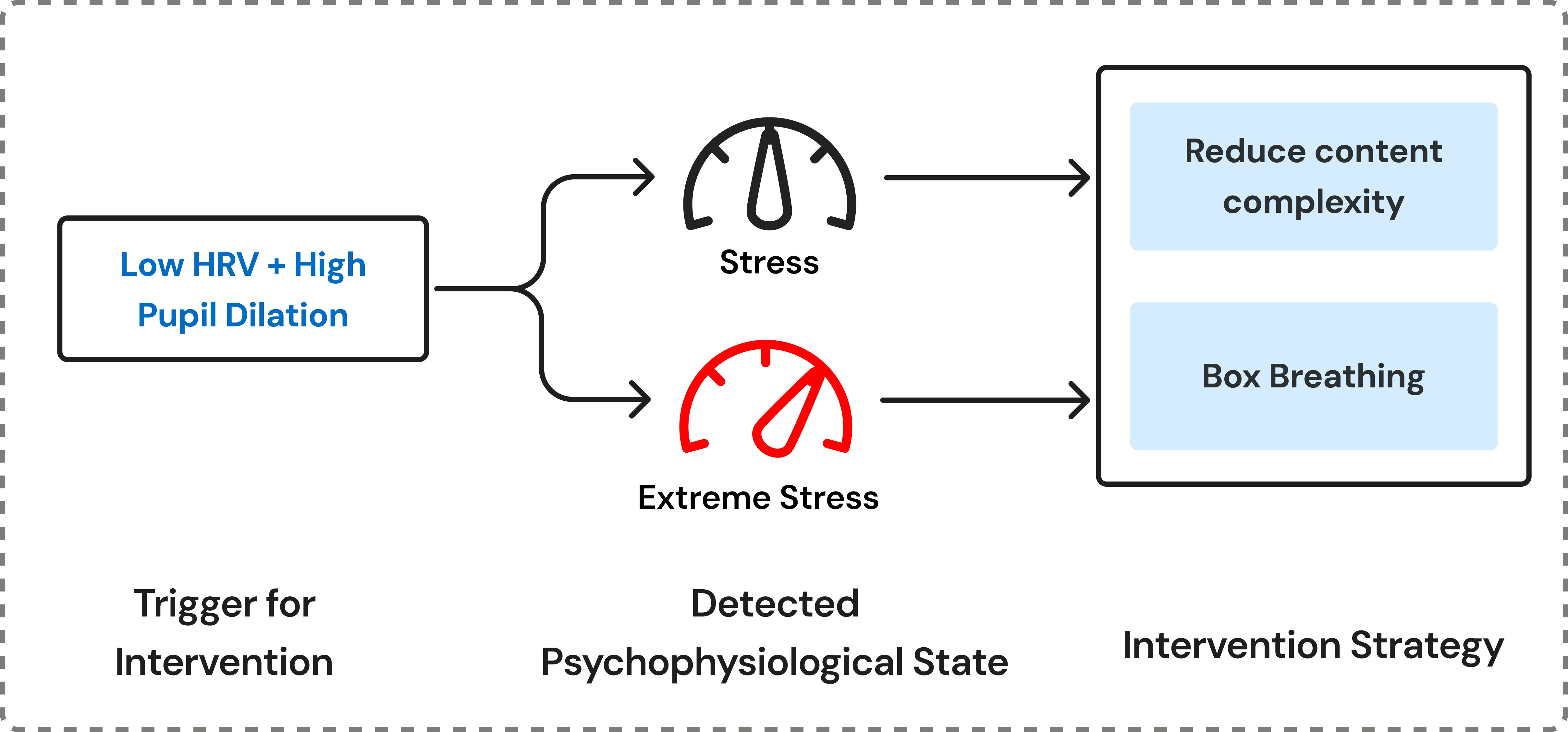} \\
        \small{(c) High Attention State Recognition} & 
        \small{(d) Stress Regulation} \\
        \\
        \includegraphics[width=0.47\textwidth]{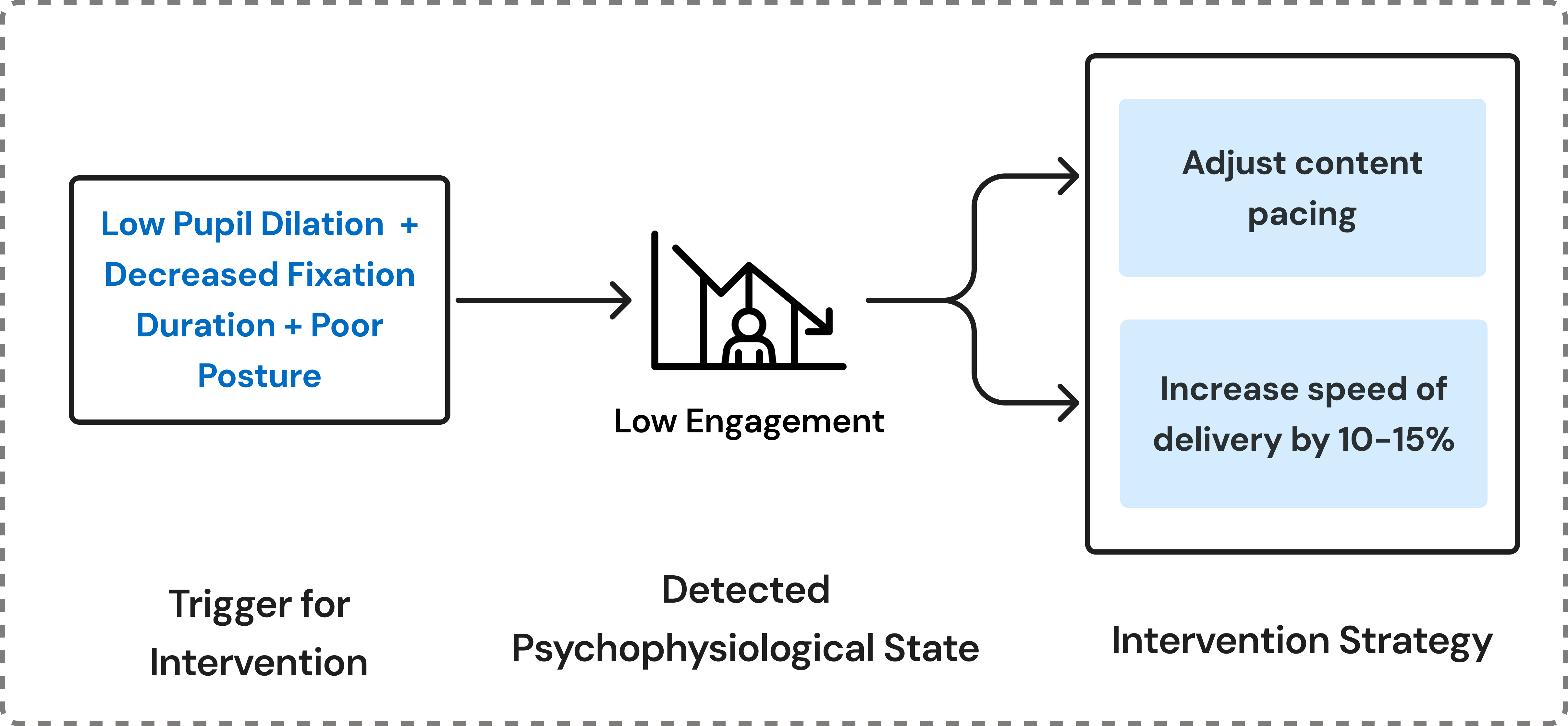} &
        \includegraphics[width=0.47\textwidth]{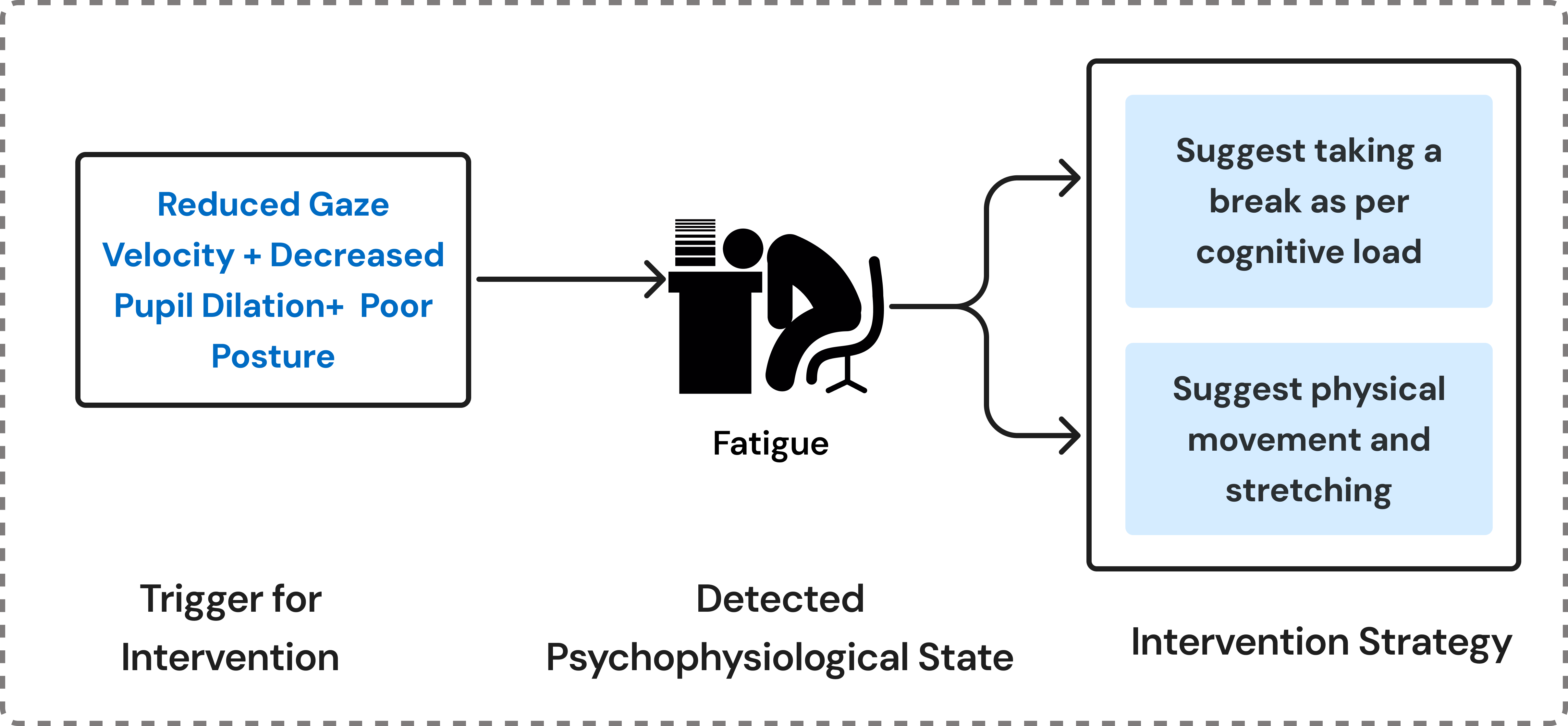} \\
        \small{(e) Low Engagement Management} & 
        \small{(f) Fatigue Detection and Response} \\
    \end{tabular}
\caption{GuideAI's biosensing-driven intervention framework linking physiological triggers to adaptive strategies. (a) High cognitive load detected via pupil dilation, elevated heart rate, and fixation duration triggers content pausing and chunking; (b) Understanding difficulties from erroneous notes initiate conceptual correction; (c) High attention states enable complexity increases; (d) Stress detection triggers graduated interventions from simplification to breathing exercises; (e) Low engagement adapts content pacing; (f) Fatigue prompts strategic breaks. This represents a subset of GuideAI's real-time pedagogical adaptations.}
    \label{fig:intervention_framework}
    \Description{Six flowcharts illustrating the logic for different intervention types:
(a) High Cognitive Load: Triggered by pupil dilation and heart rate; leads to pausing content and chunking information.
(b) Low Understanding: Triggered by erroneous notes; leads to presenting misconceptions and guiding reconciliation.
(c) High Attention: Triggered by increased fixation; leads to increasing concept complexity and speed.
(d) Stress Regulation: Triggered by low HRV; leads to reducing complexity and prompting box breathing.
(e) Low Engagement: Triggered by poor posture and low dilation; leads to adjusting pacing and delivery speed.
(f) Fatigue: Triggered by reduced gaze velocity; leads to suggesting a break or physical stretching.}
\end{figure*}

\begin{figure*}[!htp] 
    \centering
    \includegraphics[width=\textwidth]{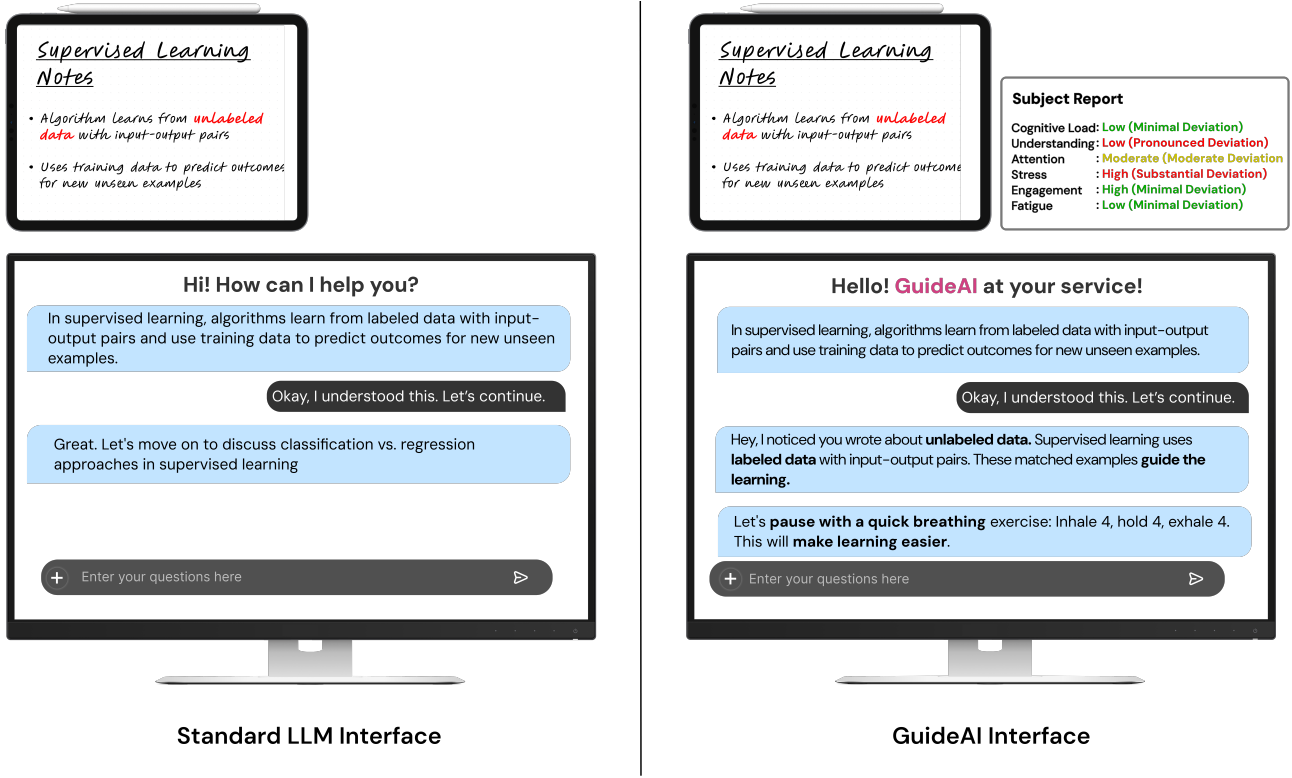} 
\caption{Comparison of standard LLM (left) vs. GuideAI (right) responses. GuideAI detects a misconception in the learner's notes (unlabeled vs. labeled data), identifies high stress levels, and delivers a tone-adaptive intervention with conceptual correction and breathing guidance to address both cognitive and affective states.}
    \label{fig:comparison}
    \Description{Side-by-side comparison of two computer screens displaying a learning chat interface about 'Supervised Learning'. 
The left screen (Standard LLM Interface) shows a basic exchange where the user acknowledges understanding and the system moves to the next topic. 
The right screen (GuideAI Interface) shows the system intervening after detecting a mistake in the user's notes. It displays a message: 'Hey, I noticed you wrote about unlabeled data... Supervised learning uses labeled data.' followed by a blue box prompt: 'Let's pause with a quick breathing exercise: Inhale 4, hold 4, exhale 4'.}
\end{figure*}

Our personalized learning system employs an intelligent intervention module that continuously analyzes the learner's state and delivers adaptive support based on multi-modal biosensing data. This adaptive architecture represents a significant departure from traditional learning systems that present standardized content regardless of the learner's cognitive and affective state.

\paragraph{Threshold Calibration and Rationale}
All intervention thresholds are defined relative to \textbf{participant-specific baselines}, obtained during a 5-minute calibration phase prior to each session (Section 5.1.1). For each physiological channel, we compute standardized deviations using a Z-score formulation:

\begin{equation}
z_t = \frac{x_t - \mu_{\text{baseline}}}{\sigma_{\text{baseline}}}
\end{equation}

where $x_t$ is the real-time signal value, and $\mu_{\text{baseline}}$, $\sigma_{\text{baseline}}$ are participant-specific baseline statistics.

We define \textbf{moderate deviation} at $|z| \geq 1.0$ and \textbf{pronounced deviation} at $|z| \geq 1.5$, consistent with prior cognitive load and stress detection literature using pupillometry and HRV-based markers. These values were further validated through pilot sessions to balance sensitivity against false-positive intervention triggering.

\paragraph{Why Semantic State Abstractions Instead of Raw Numeric Streams}
While raw physiological values (e.g., pupil diameter in millimeters or HRV metrics) are informative, directly injecting high-frequency numeric streams into LLM prompts introduces substantial noise and token inefficiency. Instead, we convert processed signals into \textbf{semantic state descriptors} (e.g., \textit{High Stress}, \textit{Moderate Cognitive Load}) that preserve interpretability while remaining stable over time.

This abstraction enables the LLM to reason over learner state using cognitively meaningful constructs, improving robustness and controllability of generated interventions without requiring the model to infer semantics from raw numerical values.

\subsubsection{Physiological State Inference Model}

\paragraph{Baseline Normalization and Signal Quantification.}
Each participant undergoes a 60-second calibration phase in a neutral resting state to compute per-sensor baseline distributions $(\mu_{\text{base},i}, \sigma_{\text{base},i})$. Real-time readings $x_{i,t}$ are standardized as $Z_{i,t} = \frac{x_{i,t} - \mu_{\text{base},i}}{\sigma_{\text{base},i}}$.

\paragraph{Dimension Scoring.}
Each of the six psychophysiological dimensions $D$ (e.g., cognitive load, attention) is computed as a weighted average of the absolute deviation scores of its constituent sensor features: 
$D_t = \frac{\sum_{i=1}^{n} w_{D,i} \cdot q_{i,t} \cdot |Z_{i,t}|}{\sum_{i=1}^{n} w_{D,i} \cdot q_{i,t}}$, 
where $w_{D,i}$ is the feature weight and $q_{i,t} \in [0,1]$ is the real-time signal quality score based on sensor confidence. Weights were derived from meta-analytic effect sizes in psychophysiology literature. For example, pupil dilation’s high weight for cognitive load reflects its documented large effect size ($d = 0.8$–$1.2$) in working memory paradigms \cite{Piquado2010}.

\paragraph{Intervention Triggering.} An intervention is triggered when a dimension score $D_t$ surpasses a predefined threshold (e.g., > 1.5 for "Moderate Deviation") for a sufficient duration (temporal persistence > 10s), a policy designed to distinguish meaningful state changes from transient noise.

\subsubsection{LLM Integration and Tone-adaptive Intervention Delivery}
The LLM serves as the primary intervention delivery mechanism, with responses enhanced by biosensory context. When biosensors detect elevated stress, the system adopts a supportive tone with patience-signaling language. During high engagement, it increases Socratic questioning. For fatigue, it becomes more energetic and concise. Tone adaptation operates across multiple linguistic dimensions—sentence complexity, encouragement frequency, explanation directness, and metaphor usage—calibrated to physiological indicators. This ensures interventions address both content modifications and optimal presentation for the learner's current state.

The system includes targeted directives based on detected states that guide response generation. For instance, high cognitive load triggers directives to simplify explanations, reduce information density, and incorporate learning aids.

An intervention is triggered only when: (i) the deviation magnitude exceeds threshold, (ii) model confidence exceeds 0.6, and (iii) deviation persists across three consecutive windows to suppress transient noise.

\subsubsection{Learning Mode–Specific Intervention Mechanisms}

GuideAI adapts interventions to the representational properties of each medium, recognizing that cognitive load and comprehension challenges manifest differently across text, image, audio, and video. This design choice follows evidence from multi-modal learning research that instructional effectiveness depends on aligning support strategies with modality-specific affordances. The four modes were selected based on learner preferences in our formative study, and the interventions were tailored accordingly.

\begin{enumerate}[leftmargin=*]
    \item \textbf{Text-based learning:} Interventions reduce extraneous load by dynamically restructuring content. For instance, dense text is segmented into shorter paragraphs, converted into bullet lists, or supplemented with reflective prompts. Example: \emph{``Let me organize this information differently to make the relationships clearer: First, the definition of sampling distribution...''}
    
    \item \textbf{Image-based learning:} Complex visuals are progressively disclosed, annotation density is adapted to cognitive capacity, and gaze-contingent highlighting draws attention to critical regions. Example: \emph{``Looking at this visualization of a sampling distribution, notice how the shape changes as sample size increases. I'll present the sequence of images in a more structured order to help clarify the relationship.''}
    
    \item \textbf{Audio-based learning:} Speech pacing and intonation are modulated in response to overload signals. Strategic pauses and summarizations aid temporal integration, while guided breathing cues are inserted during elevated stress. Example: \emph{``I notice this concept might be challenging. I'll break this down into smaller components and explain each part more deliberately.''}
    
    \item \textbf{Video-based learning:} Playback is adaptively controlled, with pauses introduced at points of detected confusion, clarifying questions embedded, and captions or overlays added for emphasis. Example: [System automatically pauses video] \emph{``Let's pause here to ensure this critical concept is clear. This segment explains a fundamental principle we'll build upon.''}
\end{enumerate}

These modality-specific mechanisms illustrate how GuideAI leverages the affordances of each medium to scaffold comprehension, regulate pacing, and maintain learner engagement.

\subsubsection{Adaptive Interventions}

GuideAI employs a tiered adaptive intervention framework across four categories: \emph{cognitive–attentional}, \emph{physiological}, \emph{comprehension-oriented}, and \emph{challenge-enhancement}. This structure draws on cognitive load theory, self-regulation research, and the principle of keeping learners within their zone of proximal development (ZPD).

\begin{enumerate}[leftmargin=*]

\item \textbf{Cognitive and Attentional Interventions:}
For moderate cognitive load, GuideAI restructures content with clearer structure or contextual cues (\emph{“Let me highlight the key relationships.”}). Under critical overload, it distills material to core ideas. GuideAI counters attention lapses with curiosity prompts (\emph{“What if we doubled the sample size?”}) or short physical resets (\emph{“Take a quick stretch before we continue.”}).

\item \textbf{Physiological Interventions:}
GuideAI responds to stress with reassurance (\emph{“Many learners find this challenging at first—let’s break it down step by step.”}) for moderate levels, and brief breathing exercises (\emph{“Let’s pause briefly—inhale 4, hold 4, exhale 6.”}) for high stress. Fatigue triggers shorter segments or breaks depending on severity.

\item \textbf{Comprehension-oriented Interventions:}
When understanding is low, GuideAI provides adaptive scaffolding. Moderate gaps trigger multiple explanation styles—formal, intuitive, or comparative. Critical gaps prompt first-principles reconstruction, moving from concrete examples to abstract concepts (\emph{“Let’s rebuild this from a tangible example before the formal definition.”}).

\item \textbf{Challenge-enhancement Interventions:}
Under-challenge states (low load, high engagement) trigger increased task complexity to sustain engagement within the ZPD. Moderate cases reduce scaffolding or add advanced applications; substantial ones extend to theoretical or synthesis prompts (\emph{“How might this apply to stratified sampling in heterogeneous populations?”}).

\item \textbf{Implicit vs. Explicit Framing:}
GuideAI adjusts feedback style based on context. Implicit adaptations (e.g., simplifying explanations) are used for moderate deviations, while explicit acknowledgments of stress or fatigue are reserved for persistent or multimodally confirmed states. This ensures sensitivity to learner differences while supporting metacognitive awareness.

\end{enumerate}

\section{Study Design and Evaluation}

\subsection{Procedure}
To evaluate GuideAI's effectiveness, we conducted a controlled comparison study with 25 participants (13 male, 12 female; age range 18-42, mean 24.7 years). We employed a within-subjects design with counterbalanced order, wherein each participant interacted with both the GuideAI system and a non-personalized control condition utilizing a standard LLM.\\

This study is intended as a preliminary validation of GuideAI under controlled conditions, focusing on STEM-oriented learning tasks where outcomes can be assessed more objectively. Generalization to more subjective domains is explored through hypotheses and discussion rather than empirical evaluation in this study.

\subsubsection{Participant Selection and Preparation}
Participants were carefully screened to ensure they had no prior knowledge of the study topics. The sample included diverse educational backgrounds across undergraduate, master's, and doctoral programs. Each participant provided written informed consent for biosensing data collection and analysis, with all data anonymized to protect privacy. Topics were selected from diverse domains including statistics (probability vs. sampling distributions), machine learning (supervised vs. unsupervised learning), and biology (innate vs. adaptive immunity) \cite{Wang2024, Hu2025, QiangZhang2024}. Prior to each session, participants underwent a calibration process establishing baseline physiological metrics, including eye-tracker calibration, ideal posture positioning, and heart rate/HRV baseline measurements. These baselines served as reference points for adaptive thresholds in the GuideAI condition.

\subsubsection{Study Conditions}
We developed a custom web-based interface, ensuring a consistent user experience across conditions. Based on our formative study, the system utilized OpenAI's models (\texttt{gpt-4o-mini} for dialogue, \texttt{gpt-4o} for vision, and \texttt{gpt-4o-mini-tts} for audio) for their balance of quality and latency. The interface design was identical in both conditions to isolate the effects of the adaptive interventions.

Participants each experienced two conditions in a counterbalanced, within-subjects design. To control for content variability, topic difficulty was matched, and sessions were spaced by at least 24 hours to mitigate fatigue. The two conditions were:

\begin{enumerate}[leftmargin=*]
    \item \textbf{{Control (Non-adaptive Baseline):}} Participants used a non-personalized LLM-based learning system with an identical sensor configuration to the experimental condition but without adaptive interventions or content personalization.

    \item \textbf{{Intervention (GuideAI):}} Participants used the fully-adaptive system, where content, pacing, and tone were dynamically personalized in real-time based on the multi-modal analysis of their physiological and behavioral signals.
\end{enumerate}

This design allows us to test our primary hypothesis by measuring the holistic effect of our integrated, biosensor-aware support system against a rigorously controlled and representative baseline.

\subsubsection{Learning Modes}
Each condition consisted of four sequential learning modes designed to engage different learning modalities, which were selected based on the preferences identified in our formative study:

\begin{enumerate}[leftmargin=*]
   \item \textbf{Text-based Learning}: Participants studied structured text content on their assigned topic.

   \item \textbf{Image-based Learning}: Participants engaged with topic-appropriate visual content, where text explanations were supplemented with illustrative images and diagrams. The content was specifically selected to benefit from visual reinforcement, for example, statistical distribution graphs for probability concepts. Each visual element included detailed explanations of its components.

   \item \textbf{Audio-based Learning}: Participants engaged in audio-based dialogue with the LLM, asking questions and receiving spoken responses about the topic, creating a more interactive learning experience.

   \item \textbf{Video-based Learning}: Participants watched an educational video on the topic, with the ability to ask the system clarifying questions about the video content.
\end{enumerate}

\subsubsection{Assessment Methods}
After each learning mode, participants completed:

\begin{enumerate}[leftmargin=*]
   \item \textbf{Cognitive Load Assessment}: NASA Task Load Index (NASA-TLX) \cite{Hart1988} on a 7-point Likert scale measuring perceived mental demand, effort, temporal demand, physical demand, performance and frustration \cite{Likert1932}.

    \item \textbf{Knowledge Assessment}: Multiple-choice questions testing factual recall and conceptual understanding through application-based problems, with similar difficulty across all topics.

   \item \textbf{Subjective Evaluation}:  Participants rated their experience on 7-point Likert scales covering perceived focus, adaptation to learning pace, quality of examples, and overall satisfaction.
\end{enumerate}
For the GuideAI condition, participants additionally evaluated the quality and appropriateness of interventions received. All participants provided qualitative feedback on both systems at the conclusion of each learning session.

\subsection{Findings}

Our evaluation demonstrates that GuideAI significantly improves both learning outcomes and cognitive experience compared to the non-adaptive control condition. Results are reported across three dimensions: (1) cognitive load and user experience, (2) learning performance, and (3) subjective evaluations.

\subsubsection{Cognitive Load and User Experience}

\begin{figure}[ht]
    \centering
    \includegraphics[width=\columnwidth]{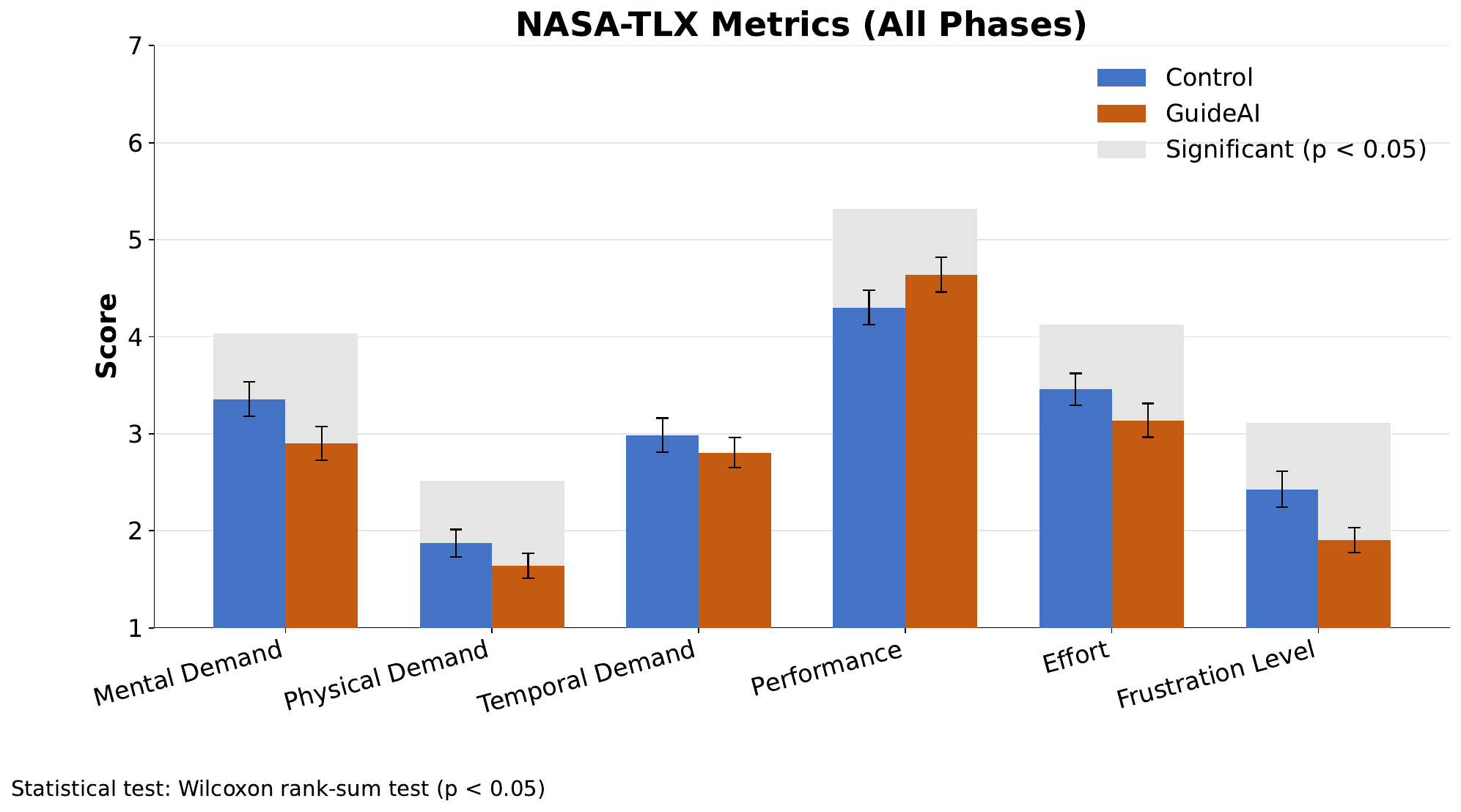}
    \caption{Comparison of NASA-TLX metrics between control and GuideAI conditions across all learning modes. Lower scores are better for all metrics except Performance, where higher scores indicate better perceived success. Gray highlighting indicates statistically significant differences (p < 0.05) using Wilcoxon rank-sum tests \cite{Rey2011}, which were used due to the non-normal distribution and paired nature of the data.}
    \label{fig:nasa_tlx_overall}
    \Description{Clustered bar chart comparing NASA-TLX scores between Control (blue) and GuideAI (orange) conditions. 
GuideAI shows lower (better) scores for Mental Demand, Physical Demand, Temporal Demand, Effort, and Frustration Level. 
GuideAI shows higher (better) scores for Performance. 
Gray shaded backgrounds indicate statistically significant differences (p < 0.05) for all metrics except Physical Demand.}
\end{figure}

NASA-TLX analysis revealed consistent reductions in cognitive load across all modalities with GuideAI (Figure~\ref{fig:nasa_tlx_overall}). Participants reported lower mental demand ($\Delta = 0.49$, $p < 0.05$), physical demand ($\Delta = 0.23$, $p < 0.05$), effort ($\Delta = 0.28$, $p < 0.05$), and frustration ($\Delta = 0.54$, $p < 0.05$), alongside higher perceived performance ($\Delta = 0.34$, $p < 0.05$). These results indicate reduced cognitive burden and improved self-assessed learning efficacy.

Learning mode-specific analysis revealed how GuideAI influenced cognitive load across different learning modalities (Figure \ref{fig:nasa_tlx_all_phases}):

\begin{figure*}[!htp]
    \centering
    \begin{subfigure}[b]{0.48\textwidth}
        \centering
        \includegraphics[width=\textwidth]{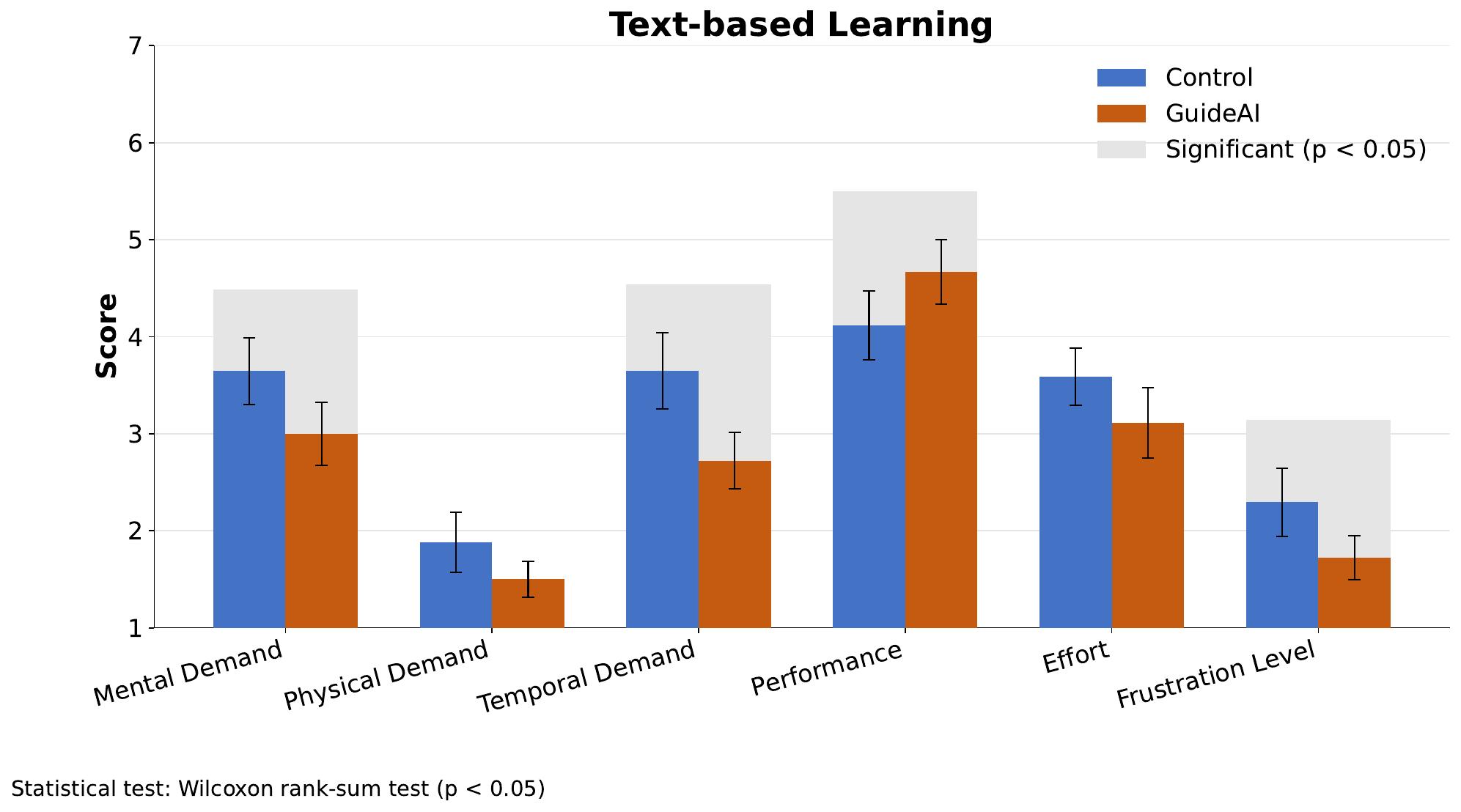}
        \caption{}
        \label{fig:nasa_tlx_phase1}
    \end{subfigure}
    \begin{subfigure}[b]{0.48\textwidth}
        \centering
        \includegraphics[width=\textwidth]{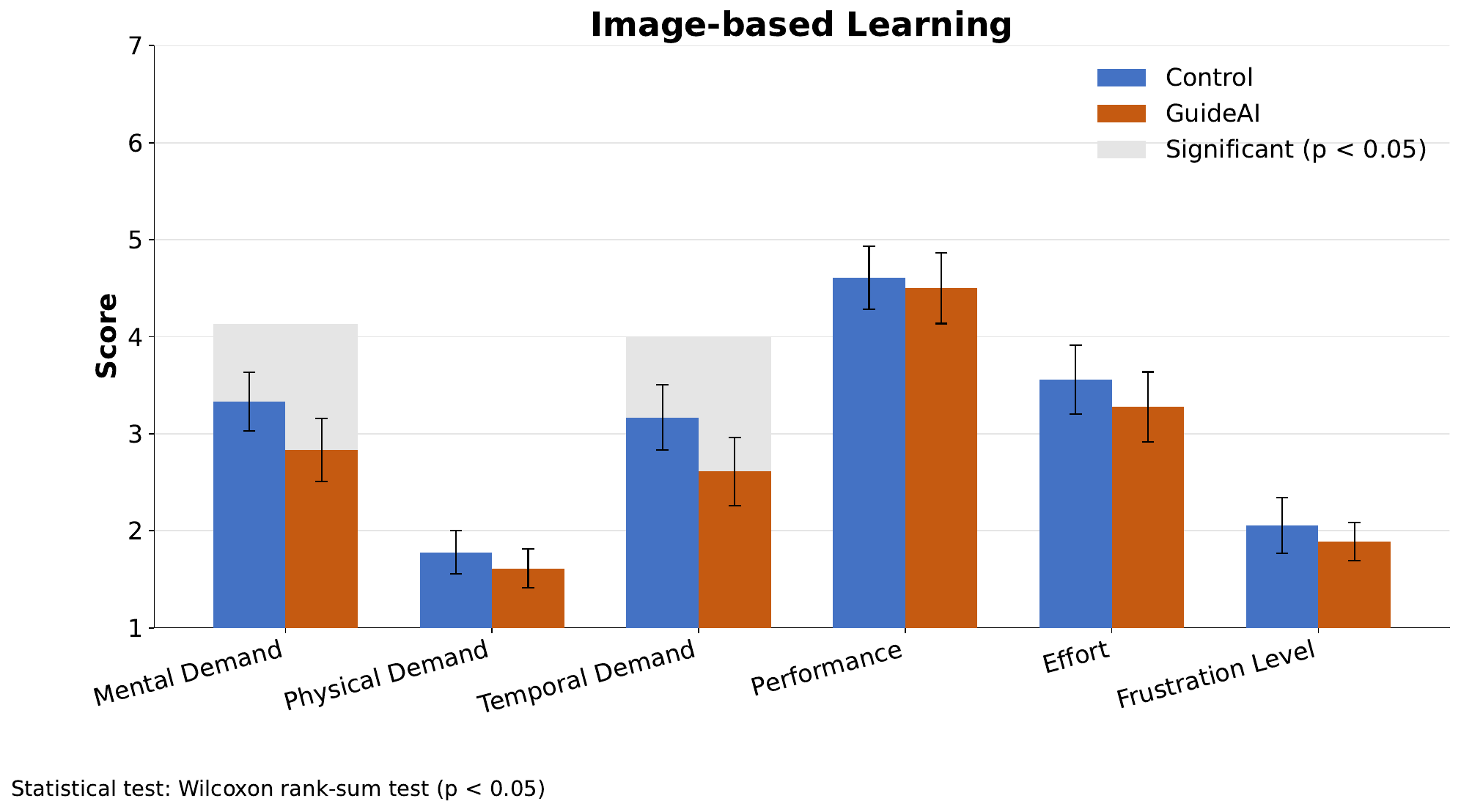}
        \caption{}
        \label{fig:nasa_tlx_phase2}
    \end{subfigure}

    \begin{subfigure}[b]{0.48\textwidth}
        \centering
        \includegraphics[width=\textwidth]{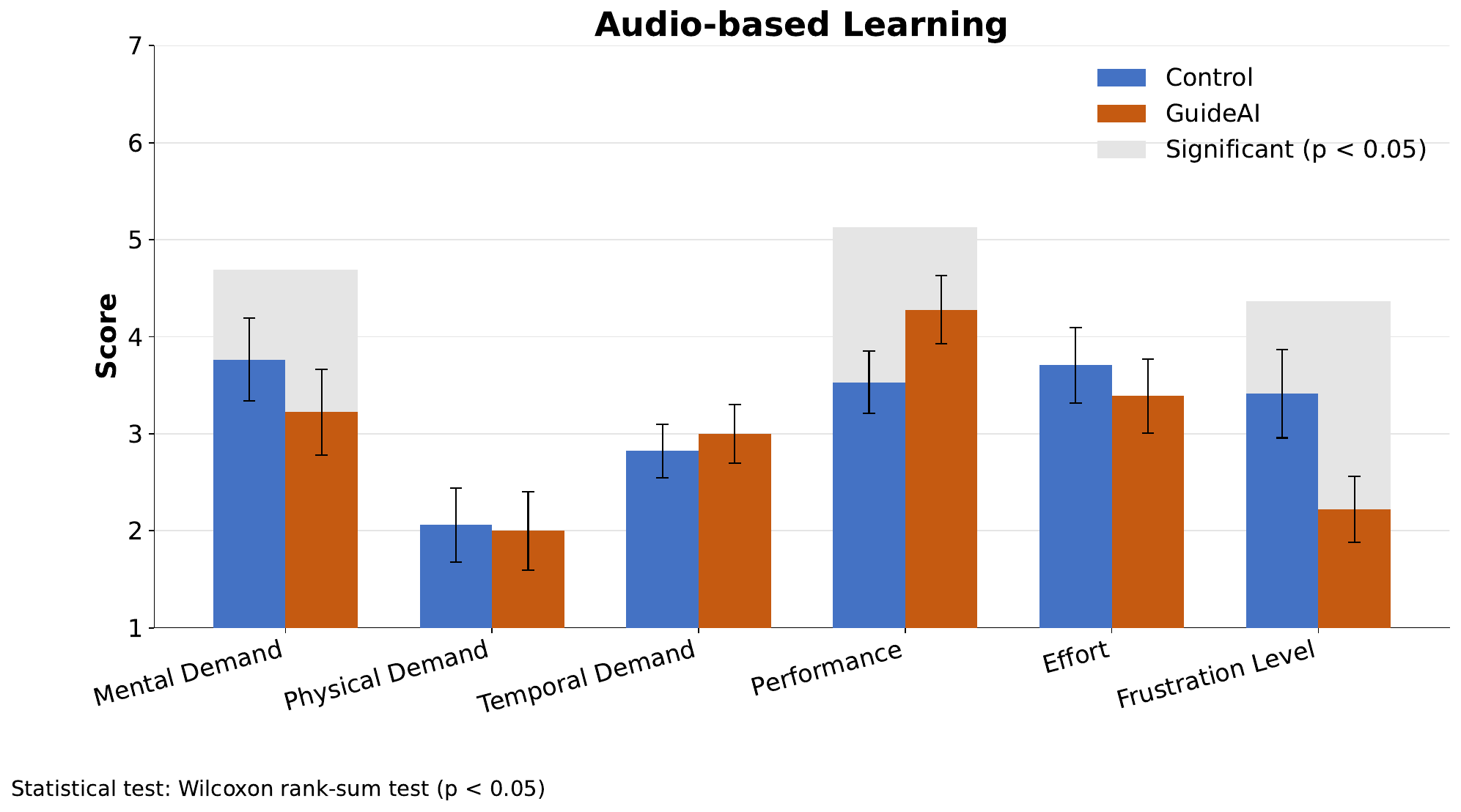}
        \caption{}
        \label{fig:nasa_tlx_phase3}
    \end{subfigure}
    \begin{subfigure}[b]{0.48\textwidth}
        \centering
        \includegraphics[width=\textwidth]{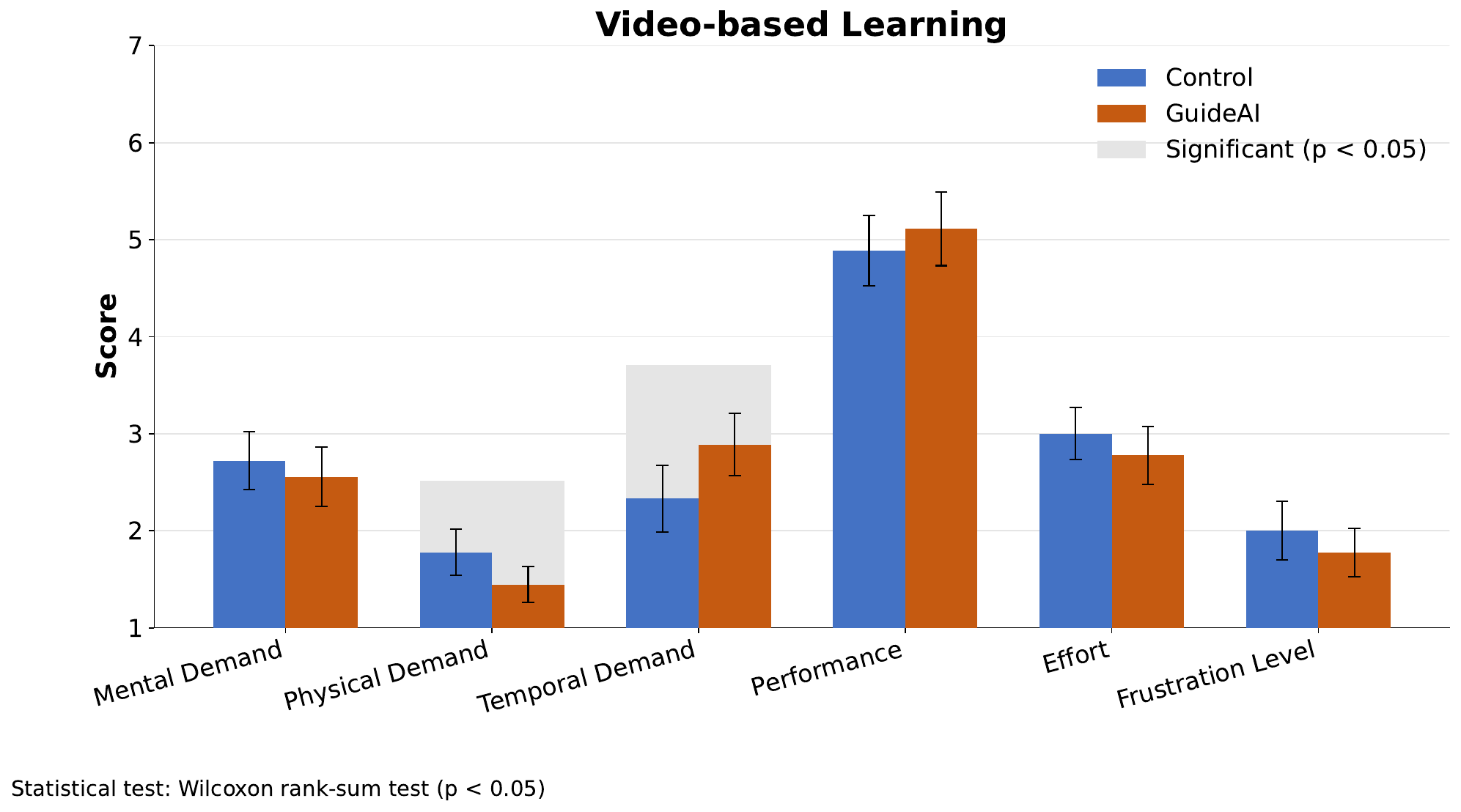}
        \caption{}
        \label{fig:nasa_tlx_phase4}
    \end{subfigure}
    
\caption{NASA-TLX metrics across learning modes comparing GuideAI vs. control. (a) Text-based learning showed significant reductions in mental demand, temporal demand, and frustration; (b) Image-based learning reduced mental and temporal demand via eye-tracking adaptations; (c) Audio-based learning showed the largest performance improvements with reduced mental demand and frustration; (d) Video-based learning showed lower physical demand but higher temporal demand due to adaptive pacing. Gray shading indicates significance (p < 0.05, Wilcoxon rank-sum). Lower is better except Performance (higher is better).}
    \label{fig:nasa_tlx_all_phases}
    \Description{Four bar charts showing NASA-TLX metrics separated by learning mode:
(a) Text-based Learning: GuideAI significantly reduces Mental Demand, Temporal Demand, and Frustration compared to Control.
(b) Image-based Learning: GuideAI significantly reduces Mental Demand and Temporal Demand.
(c) Audio-based Learning: GuideAI shows significantly higher Performance and lower Frustration.
(d) Video-based Learning: GuideAI shows significantly lower Physical Demand but higher Temporal Demand.}
\end{figure*}

\begin{itemize}
    \item \textbf{Text-based:} Largest reduction in cognitive load, particularly in mental ($\Delta = 0.65$, $p < 0.05$) and temporal demand ($\Delta = 0.96$, $p < 0.05$), suggesting strong alignment between pacing interventions and reading-intensive content (Figure \ref{fig:nasa_tlx_all_phases}a).
    \item \textbf{Image-based:} Significant decreases in mental ($\Delta = 0.51$, $p < 0.05$) and temporal demand ($\Delta = 0.59$, $p < 0.05$); performance gains were moderate (Figure \ref{fig:nasa_tlx_all_phases}b).
    \item \textbf{Audio-based:} Higher perceived performance ($\Delta = 0.77$, $p < 0.05$) and reduced frustration ($\Delta = 1.14$, $p < 0.05$), indicating improved engagement through adaptive dialogue pacing (Figure \ref{fig:nasa_tlx_all_phases}c).
    \item \textbf{Video-based:} Lower physical demand ($\Delta = 0.33$, $p < 0.05$) but higher temporal demand ($\Delta = 0.60$, $p < 0.05$), likely reflecting real-time playback adjustments synchronized to cognitive state (Figure \ref{fig:nasa_tlx_all_phases}d).
\end{itemize}

Participants' qualitative feedback reinforced these findings:

\begin{quote}
\textit{``The interventions were helpful as they were monitoring stress, focus, etc. and trying to reduce the complexity of concepts on the go, which helped in understanding things in a simpler manner.''} --- P7
\end{quote}

\begin{quote}
\textit{``I genuinely appreciated how it slowed down when I was getting overwhelmed and offered simpler explanations. The personalized pacing made a noticeable difference in how much information I was able to retain.''} --- P20
\end{quote}

\subsubsection{Learning Performance}
GuideAI produced significant learning gains across modalities for both problem-solving and recall tasks.

\begin{figure*}[t]
    \centering
    \begin{subfigure}[b]{0.48\textwidth}
        \centering
        \includegraphics[width=\textwidth]{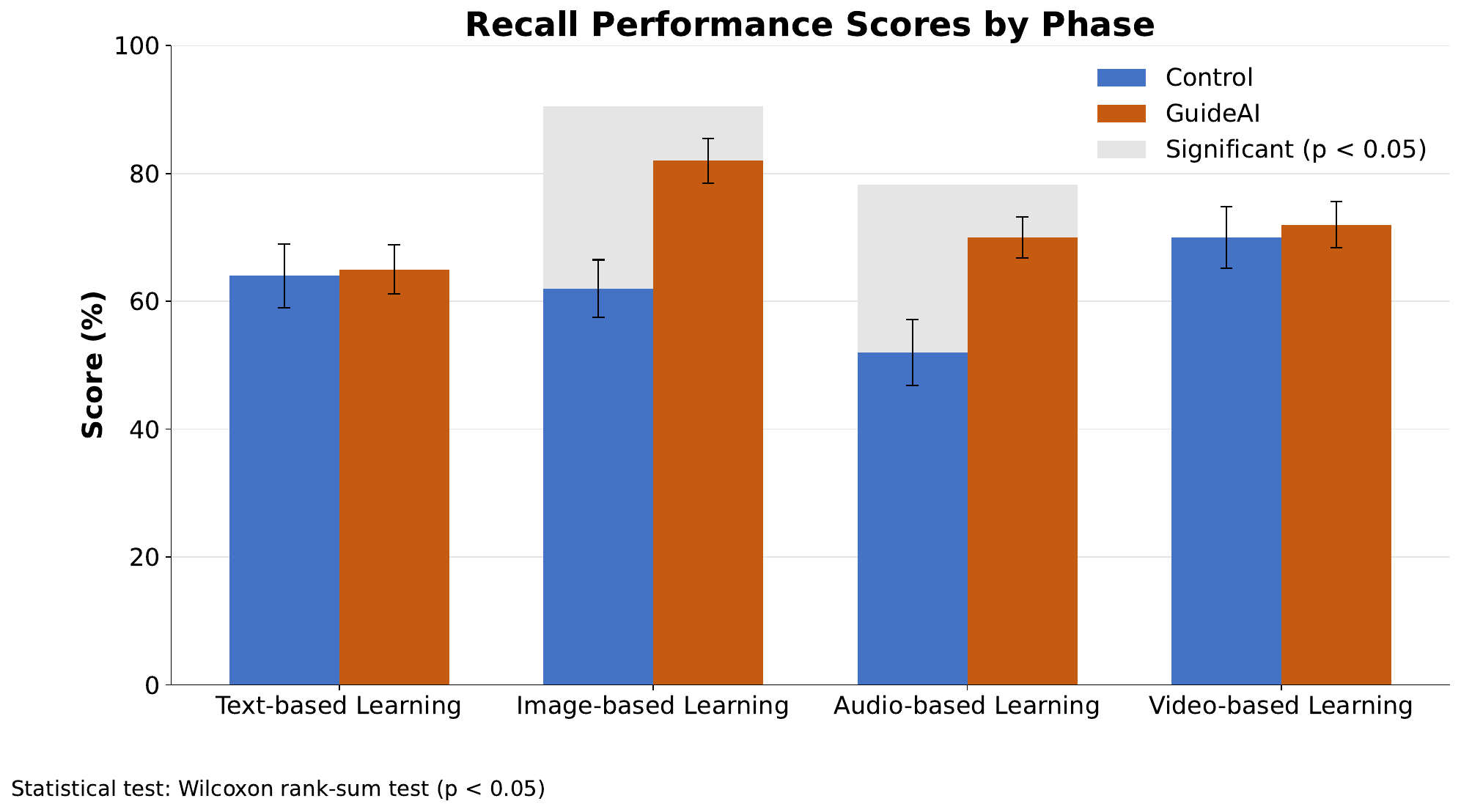}
        \caption{Recall performance}
        \label{fig:recall_scores}
    \end{subfigure}
    \hfill
    \begin{subfigure}[b]{0.48\textwidth}
        \centering
        \includegraphics[width=\textwidth]{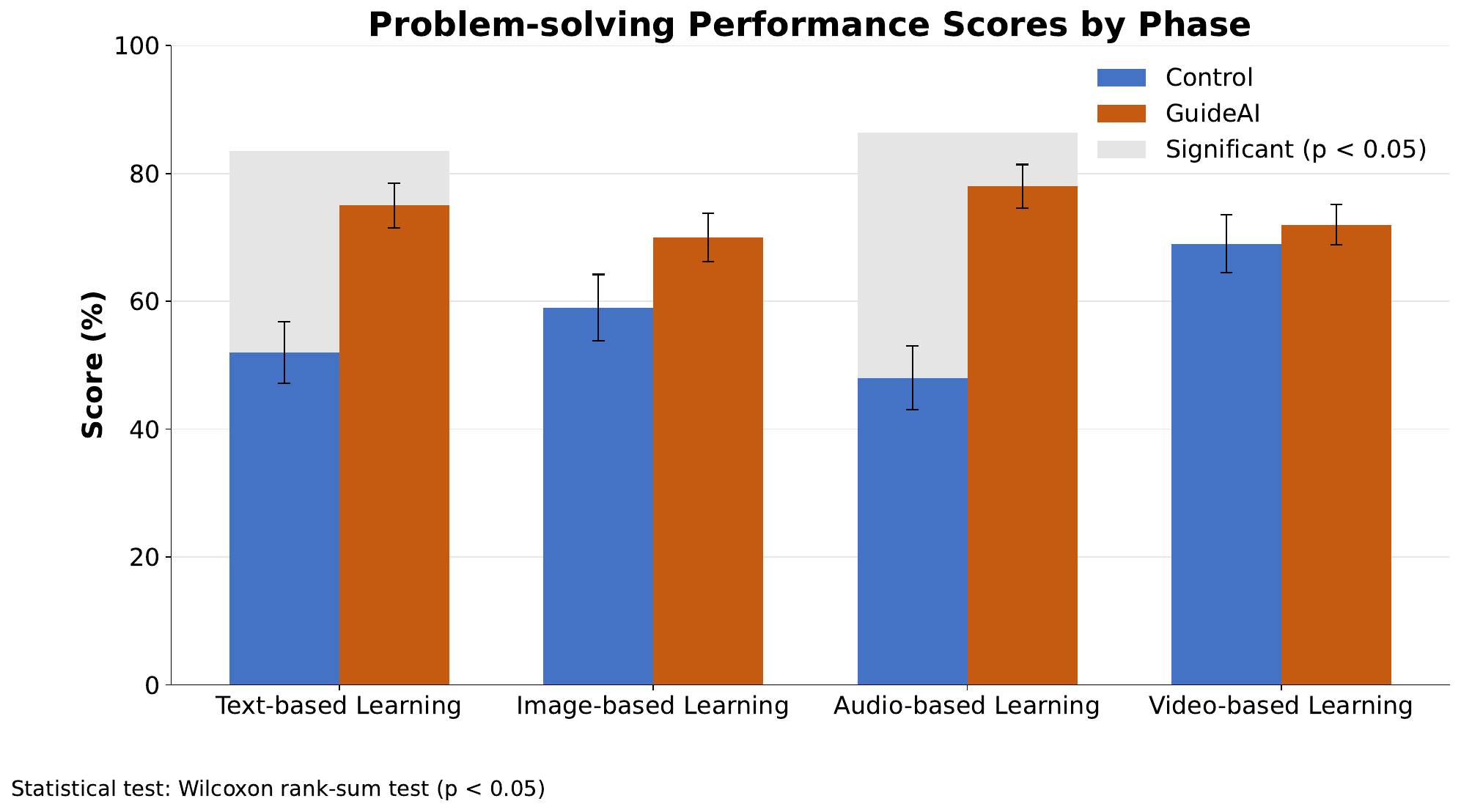}
        \caption{Problem-solving performance}
        \label{fig:problem_solving_scores}
    \end{subfigure}
\caption{Learning performance comparing Control vs. GuideAI. (a) Recall performance with significant improvements in image-based and video-based learning. (b) Problem-solving performance with significant gains in image-based and audio-based learning. GuideAI outperformed control across all modalities. Error bars: standard error. Gray shading: p < 0.05 (Wilcoxon rank-sum).}
    \label{fig:performance_combined}
    \Description{Two bar charts comparing learning outcomes. 
(a) Recall Performance: GuideAI scores are higher than Control across all modes, with statistically significant improvements marked for Image-based and Audio-based learning. 
(b) Problem-solving Performance: GuideAI scores are higher across all modes, with statistically significant improvements for Text-based, Image-based, and Audio-based learning. The largest gap is seen in Image-based learning problem solving.}
\end{figure*}

\begin{itemize}
    \item \textbf{Problem-solving:} Text-based 75\% vs 52\%, Image-based 78\% vs 48\%, Audio-based 70\% vs 59\% ($p < 0.05$); Video-based 72\% vs 68\% (n.s.). The largest gain ($\Delta = 30$ percentage points) occurred in image-based learning (Figure~\ref{fig:problem_solving_scores}).
    \item \textbf{Recall:} Image-based 82\% vs 62\%, Audio-based 70\% vs 52\% ($p < 0.05$); Text-based 65\% vs 64\% and Video-based 72\% vs 70\% (Figure~\ref{fig:recall_scores}).
\end{itemize}

Performance improvements aligned with reduced cognitive load: for instance, in image-based learning, higher recall corresponded with lower mental ($\Delta = 0.51$) and temporal demand ($\Delta = 0.59$). Audio-based gains in problem-solving paralleled reduced frustration ($\Delta = 1.14$) and increased perceived performance ($\Delta = 0.77$).
Across all modalities, GuideAI yielded a mean improvement of 16.5 percentage points in problem-solving and 10.3 percentage points in recall, indicating stronger support for deep conceptual integration than rote memorization. These findings validate that biosensor-driven adaptive interventions enhance comprehension by reducing cognitive burden and aligning feedback with learner state.

\subsubsection{Subjective Evaluation}

\begin{table*}[t]
  \centering
  \caption{Comparison of Control and GuideAI Groups on Likert Scale Responses ranging from 1 (Strongly Disagree) to 7 (Strongly Agree)}
  \label{tab:subjective_evaluation}
  \begin{tabular}{p{0.4\textwidth}|c|c|c|c|c}
    \hline
    \textbf{Question} & \textbf{Control} & \textbf{GuideAI} & \textbf{U-} & \textbf{p-} & \textbf{Effect} \\
    & \textbf{Mean/SD} & \textbf{Mean/SD} & \textbf{Statistic} & \textbf{value} & \textbf{Size} \\
    \hline
    I found the learning task to be challenging using this learning system & 4.65 / 1.22 & 4.30 / 1.18 & 159.0 & 0.1672 & 0.29 \\
    \hline
    I felt frustrated or overwhelmed while using the learning system & 4.10 / 1.32 & 3.75 / 1.28 & 165.5 & 0.2481 & 0.27 \\
    \hline
    I was able to stay focused throughout the session with the learning system & 4.85 / 1.34 & \textbf{5.60 / 1.25} & 127.5 & \textbf{0.0418} & -0.48 \\
    \hline
    I enjoyed the learning task with the learning system & 5.20 / 1.30 & 5.45 / 1.38 & 170.0 & 0.3621 & -0.19 \\
    \hline
    The examples provided by the learning system were tailored to me and easy to follow & 4.90 / 1.26 & \textbf{5.55 / 1.32} & 124.0 & \textbf{0.0362} & -0.45 \\
    \hline
    The learning system adapted well to my learning pace & 4.65 / 1.22 & \textbf{5.40 / 1.28} & 123.0 & \textbf{0.0322} & -0.47 \\
    \hline
    I feel I understood the topic better after using this learning system & 4.75 / 1.18 & \textbf{5.50 / 1.16} & 126.0 & \textbf{0.0402} & -0.49 \\
    \hline
  \end{tabular}
  \smallskip
  \flushleft
  \small \textit{Note: Statistical analysis includes the U-statistic, p-value from the Wilcoxon rank-sum tests, and effect size (Cohen's d). Negative effect sizes indicate that GuideAI condition had a higher mean value than the Control condition.}
\end{table*}

Participants’ subjective ratings (Table~\ref{tab:subjective_evaluation}) corroborated the quantitative results. Four dimensions showed statistically significant improvements in the GuideAI condition:

\begin{itemize}
    \item \textbf{Focus and Attention:} $M = 5.60$, $SD = 1.25$ vs Control $M = 4.85$, $SD = 1.34$, $p = 0.0418$.
    \item \textbf{Content Personalization:} $M = 5.55$, $SD = 1.32$ vs Control $M = 4.90$, $SD = 1.26$, $p = 0.0362$.
    \item \textbf{Adaptive Pacing:} $M = 5.40$, $SD = 1.28$ vs Control $M = 4.65$, $SD = 1.22$, $p = 0.0322$.
    \item \textbf{Perceived Learning:} $M = 5.50$, $SD = 1.16$ vs Control $M = 4.75$, $SD = 1.18$, $p = 0.0402$.
\end{itemize}

Positive but non-significant trends were observed for task challenge ($p = 0.17$) and frustration ($p = 0.25$). Overall enjoyment was comparable across conditions (GuideAI: $M = 5.45$, $SD = 1.38$; Control: $M = 5.20$, $SD = 1.30$).
Qualitative feedback further illustrated increased engagement and perceived adaptivity:

\begin{quote}
\textit{``Before using this system, I'd often give up on difficult material. The way it broke concepts down when it detected my confusion gave me capabilities I didn't have before. I could actually stay with the material instead of getting lost.''} --- P4
\end{quote}

\subsubsection{Interventions Evaluation}
\sloppy
\begin{table}[t]
  \centering
  \caption{Additional Questionnaire Responses from the GuideAI Group on Personalized Interventions}
  \label{tab:intervention_evaluation}
  \begin{tabular}{p{0.62\columnwidth}|c}
    \hline
    \textbf{Question} & \textbf{GuideAI Mean/SD} \\
    \hline
    Helped me understand the topic more clearly & 4.71 / 0.68 \\
    \hline
    Responded promptly to my needs & 5.50 / 0.75 \\
    \hline
    Provided examples that made concepts easier to grasp & 4.75 / 1.28 \\
    \hline
    Positively impacted my task performance & 5.25 / 0.58 \\
    \hline
    Personalization & 5.00 / 0.66 \\
    \hline
    Made learning experience more engaging & 5.33 / 0.47 \\
    \hline
    Adjusted based on my mistakes or struggles & 4.43 / 0.55 \\
    \hline
  \end{tabular}
  \smallskip
  \flushleft
  \small \textit{Note: Values measured on a 7-point Likert scale where higher values indicate stronger agreement.}
\end{table}

Participants in the GuideAI condition rated biosensor-driven interventions positively across all dimensions (Table~\ref{tab:intervention_evaluation}). The highest ratings were observed for \textbf{system responsiveness} ($M = 5.50$, $SD = 0.75$) and \textbf{enhanced engagement} ($M = 5.33$, $SD = 0.47$), indicating strong appreciation for GuideAI’s ability to detect and respond to cognitive-affective shifts in real time. Participants also rated \textbf{performance impact} highly ($M = 5.25$, $SD = 0.58$), suggesting that interventions effectively supported comprehension and task progress rather than functioning as peripheral features.

Personalization quality ($M = 5.00$, $SD = 0.66$) and conceptual clarity ($M = 4.71$, $SD = 0.68$) were consistently endorsed, confirming that biosensory adaptations successfully aligned with individual learning needs. The comparatively lower score for \textbf{error-based adjustments} ($M = 4.43$, $SD = 0.55$) indicates scope to refine the granularity of real-time difficulty calibration.
Qualitative feedback substantiated these findings. Participants emphasized the utility of physiological regulation cues, such as breathing or posture prompts, in managing stress and sustaining focus:

\begin{quote}
\textit{``I liked when it prompted me to take deep breaths to control my heartbeat. It helped me relax and refocus on the task when I was feeling overwhelmed by the content.''} --- P8
\end{quote}

\begin{quote}
\textit{``It was good and I liked this learning system because when I study normally, I don't notice my posture. Having something remind me helped me stay comfortable and focused for longer periods.''} --- P15
\end{quote}

\subsection{Ablation Study}

To assess the individual effectiveness of GuideAI's adaptive mechanisms, 
we performed an ablation study on the four intervention types detailed 
in Section 4.4.5's Adaptive Interventions framework. Specifically, we 
isolated: cognitive optimization (the content restructuring and 
complexity adjustments described under "Cognitive and Attentional 
Interventions"), physiological intervention (the breathing and posture 
guidance from "Physiological Interventions"), attention redirection (the 
focus management strategies from "Cognitive and Attentional 
Interventions"), and tone-adaptive communication (the affective language 
modulation described in Section 4.4.2).

\begin{figure*}[ht]
    \centering
    \includegraphics[width=0.9\textwidth]{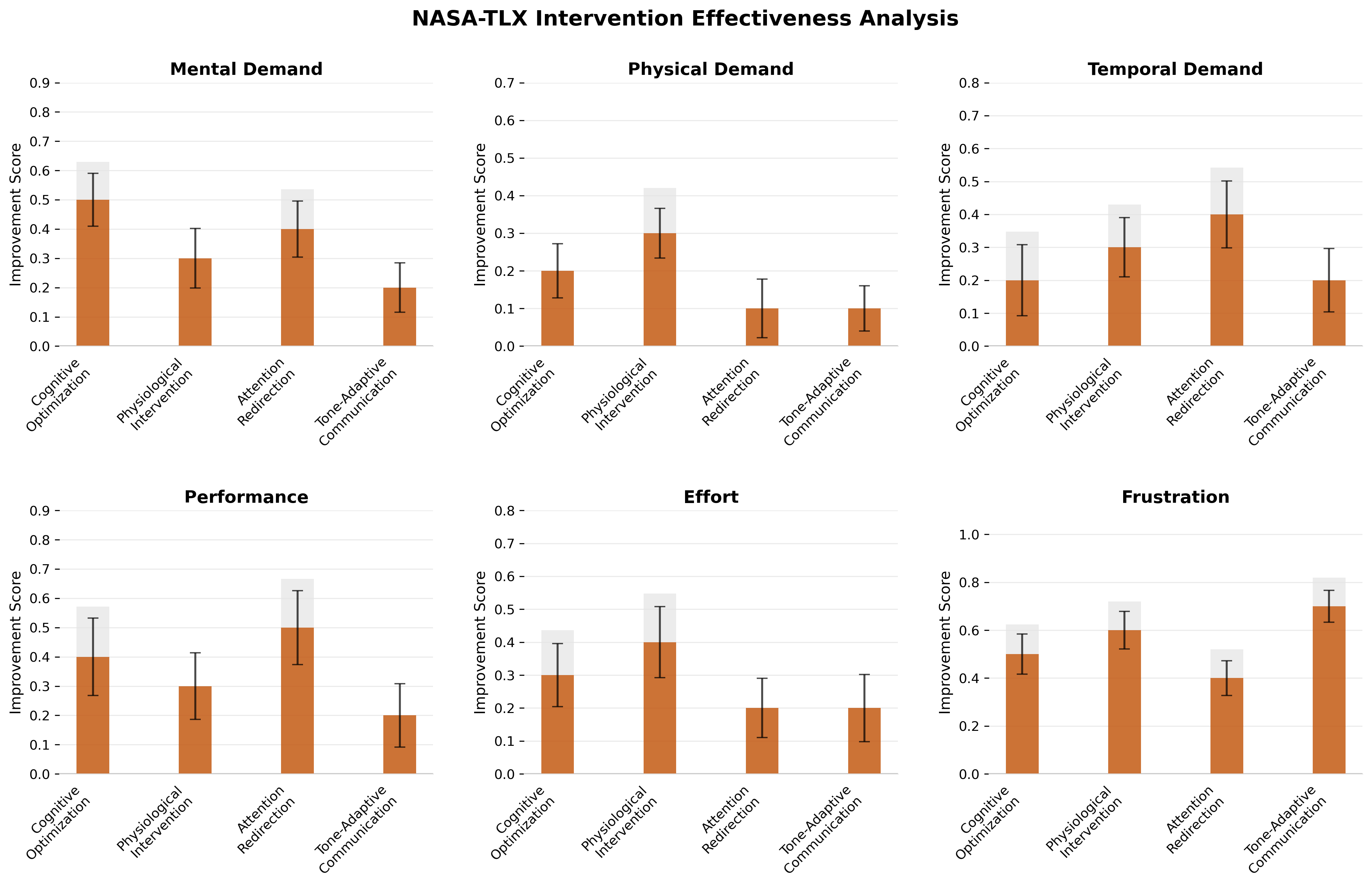}
\caption{NASA-TLX improvement scores by intervention category. Cognitive optimization shows strongest improvements; attention redirection excels at performance enhancement. Error bars: SE.}
    \label{fig:nasa_tlx_intervention_effectiveness}
    \Description{Six small bar charts displaying the 'Improvement Score' (reduction in load or increase in performance) for four intervention types: Cognitive Optimization, Physiological Intervention, Attention Redirection, and Tone-Adaptive Communication.
Cognitive Optimization shows the highest improvement scores for Mental Demand and Frustration. 
Attention Redirection shows high improvement for Temporal Demand and Performance. 
Tone-Adaptive Communication shows high improvement for Frustration.}
\end{figure*}

\subsubsection{Methodology}
For each intervention category, we calculated improvement scores by comparing NASA-TLX metrics in trials where that intervention was active versus control trials (no intervention) for the same participant. 

The improvement score is defined as the within-subject difference in NASA-TLX ratings (1-7 scale) between intervention and control trials, averaged across all six NASA-TLX dimensions. In instances where multiple interventions were triggered simultaneously (approx. 12\% of cases), the intervention associated with the highest deviation score relative to baseline was prioritized for the analysis. This approach isolates the primary driver of the state change while accounting for the system's hierarchical decision-making process.

\subsubsection{Results}
Analysis of intervention logs revealed the distribution of system actions per session: Cognitive Optimization was the most frequently triggered mechanism ($M=5.2$ times/session), reflecting the continuous need for micro-adjustments in content complexity to maintain the zone of proximal development. This was followed by Attention Redirection ($M=3.8$) and Tone-Adaptive Communication ($M=3.1$). Physiological Intervention was the least frequent ($M=1.2$), validating our hierarchical design which reserves these explicit, flow-breaking interruptions (e.g., breathing exercises) only for sustained high-stress states to preserve learning momentum.
As shown in Figure~\ref{fig:nasa_tlx_intervention_effectiveness}, cognitive optimization yielded the largest reductions in mental demand ($0.49 \pm 0.10$), effort ($0.31 \pm 0.08$), and frustration ($0.52 \pm 0.07$). Attention redirection effectively reduced temporal demand ($0.39 \pm 0.13$) and improved perceived performance ($0.49 \pm 0.13$). Physiological regulation produced moderate but consistent frustration reductions ($0.58 \pm 0.15$), while tone-adaptive communication most strongly reduced frustration ($0.72 \pm 0.11$) but showed smaller effects on other dimensions.

The analysis revealed important synergistic effects between intervention types. Combining cognitive optimization with attention redirection produced amplified benefits beyond individual effects, particularly for mental demand reduction (combined: 0.67 vs. sum of individual: 0.58).

\subsubsection{Key Findings}
Cognitive optimization emerges as the primary driver of GuideAI's effectiveness. This validates the importance of real-time content adaptation based on cognitive load indicators. Attention redirection provides crucial performance benefits for maintaining engagement, while physiological interventions serve as stress modulators. The synergistic effects demonstrate that GuideAI's multi-modal approach provides benefits beyond individual interventions in isolation, supporting the integrated system design.

\section{Discussion}
GuideAI demonstrates the potential of biosensor-augmented personalized learning to enhance educational outcomes and reduce cognitive load. Integrating real-time physiological data with adaptive content delivery enables a responsive and effective learning environment. This section discusses broader implications, limitations, and directions for future work.

\subsection{Multi-modal State-aware Learning}
Results show significant learning improvements across modalities, especially in audio- and image-based learning, suggesting modality-specific optimization. GuideAI’s adaptive mechanisms facilitate higher-order cognition, with stronger effects on problem-solving than recall. These findings extend cognitive load theory through real-time physiological monitoring, demonstrating the feasibility of state-aware systems that adapt to attention, stress, and cognitive fluctuations with minimal latency.

While our evaluation focused on STEM-oriented tasks, the underlying state-aware adaptation framework is not inherently domain-specific. Similar benefits may extend to non-technical subjects such as the humanities, social sciences, language learning, and creative disciplines, where learning often depends more on sustained attention, interpretation, and reflection than on binary notions of correctness. In these contexts, learning quality may be better captured through engagement-oriented indicators, such as reading dwell time, interaction rhythm, or reflective note-taking—rather than accuracy-based metrics, suggesting that adaptive success may be operationalized differently across domains.

\subsection{Limitations and Challenges}
Despite the promising results, several important limitations must be acknowledged:
\subsubsection{Privacy and Data Security Concerns}

The collection of biosensing data raises significant privacy considerations. While our study implemented data minimization principles and secure encryption protocols, broader deployment would require robust privacy frameworks to protect sensitive physiological information. Future implementations must address questions around data ownership, storage limitations, and user control over their physiological data. The trade-off between personalization benefits and privacy risks necessitates careful ethical consideration in system design.

\subsubsection{Technical Feasibility and Deployment Constraints}
Our current implementation requires specialized hardware including eye-tracking devices, wearable sensors, and cameras, along with calibration procedures and controlled environments that limit deployment in typical educational settings. GuideAI's effectiveness relies on accurate physiological measurements, making it vulnerable to sensor errors and calibration issues. During testing, we observed that pupillary measurements were occasionally affected by ambient lighting changes, while HRV readings showed sensitivity to minor physical movements. The complexity of multi-modal sensor integration and real-time processing demands substantial computational resources, making the system less feasible for individual learners or resource-constrained institutions. These technical and practical constraints highlight the gap between laboratory validation and real-world educational deployment.

In scenarios where dedicated physiological sensors are unavailable, future versions of GuideAI could leverage interaction-level behavioral signals as proxy indicators of learner state. Patterns such as time-on-task (dwell time), mouse cursor movement dynamics, response latency, and interaction rhythm have been shown to relate to user attention, cognitive effort, and stress. While noisier and less direct than physiological biosignals, these interaction-derived proxies enable sensor-light or sensor-free deployments and support a graceful degradation of adaptivity in resource-constrained or privacy-sensitive environments \cite{Huang2012, Gonzalo2025, Sun2014, Pingmei2016}.

\subsubsection{LLM Hallucinations and Content Reliability}
While we implemented guided prompts and structured reasoning frameworks to mitigate LLM hallucinations, the risk of generating inaccurate or misleading content remains a concern. During our study, we observed occasional instances where the LLM generated plausible but factually incorrect explanations, particularly when responding to edge case questions. This highlights the need for content verification mechanisms and domain-specific knowledge guardrails in educational applications of LLMs \cite{Huang2025}.

\subsubsection{Longitudinal Study Considerations}
Our evaluation focused on immediate learning effects within controlled sessions rather than longitudinal outcomes. The lack of long-term assessment leaves open questions about knowledge retention over time and the potential development of dependency on adaptive interventions.

\subsubsection{Intervention-specific Effectiveness}
While our system implemented various intervention types, we did not systematically evaluate their relative effectiveness. Participant feedback indicated varying preferences for different intervention strategies, with physiological regulation features (e.g., breathing guidance, posture correction) receiving particularly positive mentions, while explicit cognitive state notifications sometimes increased anxiety rather than alleviating it. More granular analysis of intervention-specific effects would help optimize future implementations.

\subsubsection{Sample Size and Scope of Validation}
The present study should be interpreted as a preliminary validation of GuideAI rather than a conclusive assessment of its effectiveness across populations. While the sample size was sufficient to observe consistent within-subject trends and statistically meaningful effects under controlled conditions, it limits the generalizability of the findings. Larger and more diverse cohorts are required to evaluate robustness across demographic groups, learning styles, and prior knowledge levels. Future work should prioritize scaled evaluations to validate external applicability and reduce sampling bias.

\subsection{Future Directions}
Building on our findings and addressing the limitations identified, we propose several promising directions for future research:

\begin{enumerate}[leftmargin=*]
    \item \textbf{Accessible Sensing Technologies.} Future work should focus on developing cost-effective and unobtrusive sensing solutions leveraging ubiquitous devices such as webcams for gaze tracking and smartwatches for HRV measurement. Such scalable alternatives could democratize biosensor-enhanced learning across diverse educational contexts.

    \item \textbf{Longitudinal Learning Effects.} Extending evaluation beyond short-term sessions to longitudinal studies can reveal sustained effects on retention, metacognitive strategy development, and learner engagement. Examining how system familiarity and trust evolve over time will clarify the durability of adaptive intervention benefits.

    \item \textbf{Learner-configurable Interventions.} Mixed qualitative feedback underscores the need for user agency in adaptive systems. Allowing learners to configure intervention frequency, type (implicit vs. explicit), and visibility could enhance comfort, mitigate over-reliance, and align adaptive strategies with individual learning preferences.

    \item {
    \textbf{Designing Transparent Adaptive Interventions.} While GuideAI internally tracks the signals and conditions that trigger adaptive interventions, this information is not currently exposed to learners in order to avoid inducing self-monitoring bias or anxiety. Future iterations should explore selective and user-controlled ways of communicating the rationale behind interventions (e.g., indicating that an adjustment was made due to signs of increased cognitive load), balancing transparency with minimal disruption to the learning experience.
    }
\end{enumerate}

\section{Ethical Considerations}
Our study's use of biosensory data followed a robust ethical framework with ethics committee approval, centered on data minimization, user control, and secure processing.
\begin{enumerate}[leftmargin=*]
    \item \textbf{Anonymization and Minimization.} All data was immediately anonymized with a participant ID. Raw sensor data was stored locally only for the session and permanently deleted after the extraction of anonymized features (e.g., mean HRV), which were retained for analysis.
    
    \item \textbf{Transparency and Consent.} Participants were fully informed about the data being collected and its purpose. A persistent on-screen indicator showed when monitoring was active, and participants could withdraw consent at any time to have their data immediately deleted.
    
    \item\textbf{Privacy and Future Directions.}To mitigate privacy risks from third-party LLM APIs, our system transmitted only anonymized, feature-extracted data (such as summary statistics or note transcriptions) rather than raw sensor streams. For real-world deployment, we strongly advocate for prioritizing on-device processing to ensure sensitive biosensory data never leaves the user's device, thus maximizing privacy and user trust.

    \item\textbf{Research Involving Human Participants and Subjects} This research involved human participants and was conducted in accordance with ethical principles and applicable regulations. The study was approved by our institution's ethics committee prior to data collection. All 25 participants provided informed consent after being fully informed about the study's purpose, procedures (including biosensory data collection via eye tracking, heart rate monitoring, posture analysis, and note-taking screenshots), potential risks, and their right to withdraw at any time without penalty.
\end{enumerate}

\section{Conclusion}
In this paper, we introduced GuideAI, a multi-modal framework that enhances LLM-driven learning by integrating real-time biosensory feedback, including eye gaze, heart rate variability, posture, and digital note-taking behavior. GuideAI dynamically adapts learning content, pacing, and interventions across diverse modalities (text-based, image-based, audio-based, and video-based) based on inferred cognitive and physiological states. Our preliminary study (N=25) demonstrated statistically significant improvements in information recall and conceptual understanding compared to standard LLM interactions, highlighting GuideAI's potential to bridge the gap between current LLM systems and the personalized, adaptive needs of learners.

While GuideAI shows promise, several avenues exist for future work. The current evaluation involved a limited number of participants and learning duration; longitudinal studies with larger, more diverse groups in real-world educational settings are needed to assess long-term effectiveness and generalizability. Exploring the integration of additional biosignals or contextual factors could further refine the system's understanding of learner states. Future iterations could also investigate the system's performance across a wider variety of learning domains and task types beyond those tested. Finally, addressing potential usability and privacy concerns associated with continuous biosensing, perhaps through on-device processing or alternative sensing methods, will be crucial for broader adoption. This work represents a step toward truly cognition-aware educational technology, adaptable to the dynamic state of individual learners.

The research adhered to principles of informed consent, privacy protection, risk minimization, and data security. All physiological and behavioral data were anonymized immediately upon collection, with personal identifiers replaced by participant codes (P1-P25). Data were encrypted and stored on secure servers with restricted access, to be retained only for the period necessary for research validation.

The research complied with the ACM Code of Ethics and Professional Conduct, The Declaration of Helsinki, and applicable local regulations governing human subjects research.

\section{GenAI Usage Disclosure}
The authors used generative AI language models (Claude by Anthropic and ChatGPT by OpenAI) solely for better phrasing and language standardization for consistency with ACM publication standards. No generative AI was used for research design, data analysis, figure generation, literature review, or writing of original content. All intellectual contributions, experimental work, and substantive writing were produced entirely by the authors, who take full responsibility for the accuracy and integrity of this work.

\begin{acks}
The authors are thankful to Harish and Bina Shah School AI \& CS and the Office of Research at Plaksha University for providing seed financial support through the Startup Research Grant Ref. No. OOR/PU-SRG/2023-24/06 for this research work.
\end{acks}

\bibliographystyle{ACM-Reference-Format}
\bibliography{references}
\end{document}